\documentclass[a4paper,usenatbib,useAMS]{mnras}
\usepackage{epsfig,amssymb,amsmath, rotating,lineno,graphicx}
\usepackage[T1]{fontenc}
\usepackage{cleveref}
\usepackage{ae,aecompl}
\usepackage{xcolor}
\usepackage{makecell}
\usepackage{subcaption}
\usepackage{caption}
\usepackage{hyperref}
\newcommand\chandra{{\it Chandra}}

\newcommand\xmm{{\it XMM-Newton}}
\newcommand\Hisaki{{\it Hisaki}}
\newcommand\IUE{{\it IUE}}
\newcommand\HST{{\it HST}}
\newcommand\GOES{{\it GOES}}

\usepackage{titlesec}
\titleformat{\subsubsection}
  {\bfseries}       
  {\thesubsubsection} 
  {1em}
  {}

\title[Jovian UV-X-ray]{A long-term spectro-temporal study of Jovian X-ray and Ultraviolet response to solar activity} 

\author[Tomer et al.]
{Megha Tomer$^{1}$, Mayukh Pahari$^{1}$, Anurag Baruah$^{1}$, Renu Malhotra$^{2}$\\
$^{1}$ Department of Physics, Indian Institute of Technology, Hyderabad, Kandi, Sangareddy 502285, India \\
$^{2}$ Lunar and Planetary Laboratory, The University of Arizona, 1629 E. University Blvd., Tucson, 85721, AZ, USA} 

\begin{document}

\pagerange{\pageref{firstpage}--\pageref{lastpage}} \pubyear{2026}

\maketitle

\label{firstpage}

\begin{abstract}
We present results from a multi-decade investigation of solar activity-driven variability in Jupiter's emissions, using solar X-ray flux and sunspot numbers as activity indicators and ultraviolet (UV) and X-ray observations from the \textit{International Ultraviolet Explorer} (\IUE\; 1978-1996) and the \textit{Chandra X-ray Observatory} (2011-2021). Analysis of 51 high-SNR UV spectra spanning two solar cycles shows that Jupiter's Ly$\alpha$ emission contains narrow and broad components, likely associated with the disk and auroral regions, respectively. The Ly$\alpha$ line and the 1330-1400~\AA continuum flux closely follow variations in solar X-ray flux and sunspot numbers throughout all phases of two consecutive solar cycles, indicating a direct connection with solar irradiation processes, including resonant scattering of solar Ly$\alpha$ photons and photoelectron-driven atmospheric excitation. In contrast, ionised UV lines such as Fe\,\textsc{ii} (1608~\AA and 1575~\AA) show no correlation with solar activity over a solar cycle, suggesting an internal or magnetospheric origin, potentially linked to Io-derived charged particles or UV/X-ray radiation.
To determine whether Jupiter's X-ray response resembles its UV response to solar activity, we analysed 29 \textit{Chandra}/HRC observations obtained during 2014-2021 and two \textit{Chandra}/ACIS observations from 2011. Significant X-ray flares are detected in both ACIS and HRC lightcurves 7-15 days after major reported coronal mass ejections (CMEs). Post-CME ACIS auroral spectra reveal a significant ($\geq 3\sigma$) Ne$^{8+}$ emission feature ($\sim$0.94-0.98~keV) near 70-80$^\circ$ latitude at Jupiter's north pole. Owing to the high ionisation energy required ($\sim$1.19~keV for the Ne \textsc{viii} to Ne \textsc{ix} bound-bound transition), this feature is unlikely to arise from local interactions, supporting CME-driven auroral excitation on Jupiter. 
\end{abstract}
\begin{keywords}
planets and satellites: aurorae -- X-rays: individual: Jupiter -- ultraviolet: planets -- Sun: coronal mass ejections (CMEs) -- techniques: spectroscopic
\end{keywords}

\section{Introduction}           
\label{sect:intro}
Jupiter emits the most intense auroral emissions in the Solar System, powered by its rapid rotation and strong magnetic field interaction with trapped charged particles (either from Solar origin or internal plasma activities). These auroral emissions have been observed across multi-wavelength regimes including radio, UV, and X-ray bands \citep{2002Natur.415..997C, 2004HEAD....8.0510E, 2004A&A...424..331B, 2006aogs....3..203B, 2006JGRA..11111225B, 2007A&A...463..761B, 2008JGRA..113.8229K, 2010JGRA..115.7102H, 2015SSRv..187...99B, 2020JGRA..12527219D, 2020JGRA..12527222D, 2022NatAs...6..442M}. Among multi-wavelength emissions, UV and X-ray regimes are particularly significant because they directly constrain the energy input, composition, and dynamics of the magnetosphere \citep{2002Natur.415..997C, 2023JGRA..12831656W}.

The Voyager UV spectrometers \citep{1979Sci...206..962S}, followed by the \IUE\, provided the first extensive spectroscopic detections of Jupiter, identifying strong H-Ly$\alpha$ and H$_{2}$ band emission together with a suite of atomic and ionic lines that trace ionospheric and magnetospheric processes \citep{1979Sci...204..979B, 1980ApJ...241L.179C}. These UV features arise predominantly from electron-impact excitation of atmospheric H and H$_{2}$ \citep{Yelle1986} and from resonant and collisional excitation of minor species originating in the Io plasma torus \citep{Bagenal1994}. As such, UV spectroscopy directly constrains precipitating-electron energies, auroral power, and the coupling between Io-supplied plasma and the magnetosphere \citep{{1996Sci...274..404C}}. Frequent monitoring in EUV/UV by subsequent missions like the Hubble Space Telescope (\HST\,) and the Japanese \Hisaki\  mission have extended these insights, linking short-term auroral morphology changes and long-term torus variability to both internal magnetospheric processes and external solar wind activities \citep{2002Natur.415..997C, 2014SSRv..184..237Y}.

While the solar irradiation of the upper Jovian atmosphere causes primary UV emission, X-ray aurorae, on the other hand, provide complex diagnostics by probing the composition of the magnetospheric plasma and its coupling to the solar wind. Theoretical studies over the past decades have established that Jovian X-ray emission originates from multiple physical mechanisms operating in both the auroral and low-latitude atmosphere. For example, X-ray emissions originating from equatorial/disk region are largely controlled by scattering and fluorescence of solar X-rays \citep{2006JGRA..111.7308C, 2005GeoRL..32.3S08B, 2022JGRA..12730971M,  2023MNRAS.521.5596W}. Early models by \citet{1995JGR...10017153C} demonstrated that energetic precipitation of highly ionised oxygen and sulfur ions from the magnetosphere into the polar atmosphere can produce soft X-ray aurorae through charge exchange, electron capture, and subsequent line emission. Later work by \citet{2003JGRA..108.1465C} further connected these auroral X-rays to magnetosphere-ionosphere coupling processes, including field-aligned acceleration, plasma transport, and reconnection-driven particle injections. In contrast, non-auroral disk X-rays are understood to arise mainly from scattering and fluorescence of incident solar X-rays in Jupiter's upper atmosphere, as modelled by \citet{2006JGRA..111.7308C}, explaining the observed correlation between disk X-ray brightness and solar activity. More recently, \citet{2013GeoRL..40.4144O} developed detailed auroral ion precipitation models for Jupiter and predicted characteristic X-ray signatures expected from precipitating heavy ions observed by \textit{JUNO}, reinforcing the view that Jovian auroral X-rays are primarily controlled by magnetospheric ion dynamics rather than direct solar forcing.

Jupiter's X-ray emissions were first detected with the \textit{Einstein} Observatory \citep{1983JGR....88.7731M} and subsequently investigated in greater detail with \textit{ROSAT}, \chandra\, and \xmm\, revealing their spatial, spectral, and temporal properties \citep{1994JGR....9914799W, 1998JGR...10320083G, 2002Natur.415.1000G,2004A&A...424..331B, 2005JGRA..110.1207E,2006JGRA..11111225B,2007A&A...463..761B, 2008JGRA..113.2202B, 2020JGRA..12527219D,2018JGRA..123.9204J}. 
Auroral X-rays are spatially concentrated in the polar zones and are spectrally complex: the soft X-ray band ($<$2 keV) is dominated by line emission produced by charge exchange of highly charged heavy ions (notably O\textsc{vii}/O\textsc{viii} features in the $\sim$0.5–0.8 keV range and putative S/C features at $\sim$0.25–0.35 keV), while emission above 2 keV is primarily a continuum from energetic-electron bremsstrahlung \citep{2005JGRA..110.1207E, 2007A&A...463..761B, 2008JGRA..113.2202B, 2022NatAs...6..442M}. Reviews and targeted campaigns \citep[e.g.,][]{2000RvGeo..38..295B, 2004A&A...424..331B, 2007A&A...463..761B} emphasise that while solar-wind compressions can trigger enhancements, internal drivers-particularly ion mass loading from Io and magnetospheric injections play a major role in powering auroral X-rays \citep{2020JGRA..12527676W}. Observationally, the northern auroral region often exhibits stronger and more persistent X-ray hot spots (e.g., the pulsating northern hot spot discovered by \chandra\, \citep{2002Natur.415.1000G}, a dominance linked to asymmetric magnetospheric topology and the preferential mapping of certain field-aligned current systems into the northern polar cap \citep{2002Natur.415.1000G, 2005JGRA..110.1207E}.

 Missions such as \IUE, \HST, and \Hisaki\ have each provided valuable insights into energetic activities of the Jovian surface and poles. For example, \citet{2022GeoRL..4997390D} analysed 14 simultaneous \chandra{} and \HST{} observations of Jupiter’s northern aurora (2016-2019) and found that the dark polar region (DPR), already known to be UV-faint, also produces almost no intrinsic X-ray emission. X-ray observations with \chandra\ and \xmm\ have revealed rich auroral dynamics, pulsations \citep{2016JGRA..121.2274D}, and ion line features, but analyses have typically focused on individual observing windows or coordinated campaigns. 
Although few individual spectral lines in the soft X-ray band have been reported \citep{2004A&A...424..331B, 2006aogs....3..203B, 2005JGRA..110.1207E, 2010JGRA..115.7102H}, their variability across longer timescales and broader significance remain comparatively less explored.  
Much of the work done in the field has been centred on short-term campaigns, isolated events, or spectra from particular epochs. To our knowledge, no comprehensive and systematic study has yet been undertaken to model the long-term evolution of Jupiter's auroral line and continuum flux across multiple solar cycles within a unified framework that links these emissions to both internal magnetospheric processes and external solar activity.

Therefore, driven by curiosity, we aim to investigate the variability of Jovian UV and auroral X-ray emissions across multiple solar cycles using archival observations. We analysed archival ultraviolet spectra from the \IUE\ (1978-1996), providing a baseline of two subsequent solar cycles for Jupiter's UV line and continuum flux. To determine the temporal variability of Jupiter’s X-ray auroral emission, we analysed the entire \textit{Chandra}/HRC catalogue (2014–2021) in a statistical study, focusing on typical and flaring behaviours of the auroral zones through light curves to derive average count rates from Jupiter's X-ray emission and quantified their statistical reliability. We also analysed the \textit{Chandra}/ACIS observations from 2011 and identified an additional emission feature near $\sim$0.94~keV which has not been reported in previous studies of Jupiter's X-ray aurora. 

To place Jupiter’s UV/X-ray activity in the context of solar variability, we systematically compared all UV and X-ray datasets with contemporaneous solar X-ray flux measurements from the Geostationary Operational Environmental Satellite (\GOES\,) in the 1.55-12.4 keV range, as well as with sunspot number indices spanning multiple solar cycles. This approach allowed us to probe the correspondence between solar activities and Jovian auroral response on long timescales. Overall, this study aims to place the multi-solar-cycle behaviour of Jovian auroral emissions within the broader framework of solar activity, magnetospheric dynamics, and transient extreme solar events.

 We outline the observations and data reductions in Section \ref{sect:obs}, while the analysis methods and results are explained in Section \ref{sect:data}. The correlation of Jupiter’s UV and X-ray emissions with solar activity is presented in Section \ref{sect:result}, and the discussion and conclusions are given in Section \ref{sect:conclusion}.

\section{Observations and data reductions}
\label{sect:obs}
\subsection{Solar Activity Proxies and Observations}
\begin{figure*}
    \centering
    \includegraphics[scale=0.47]{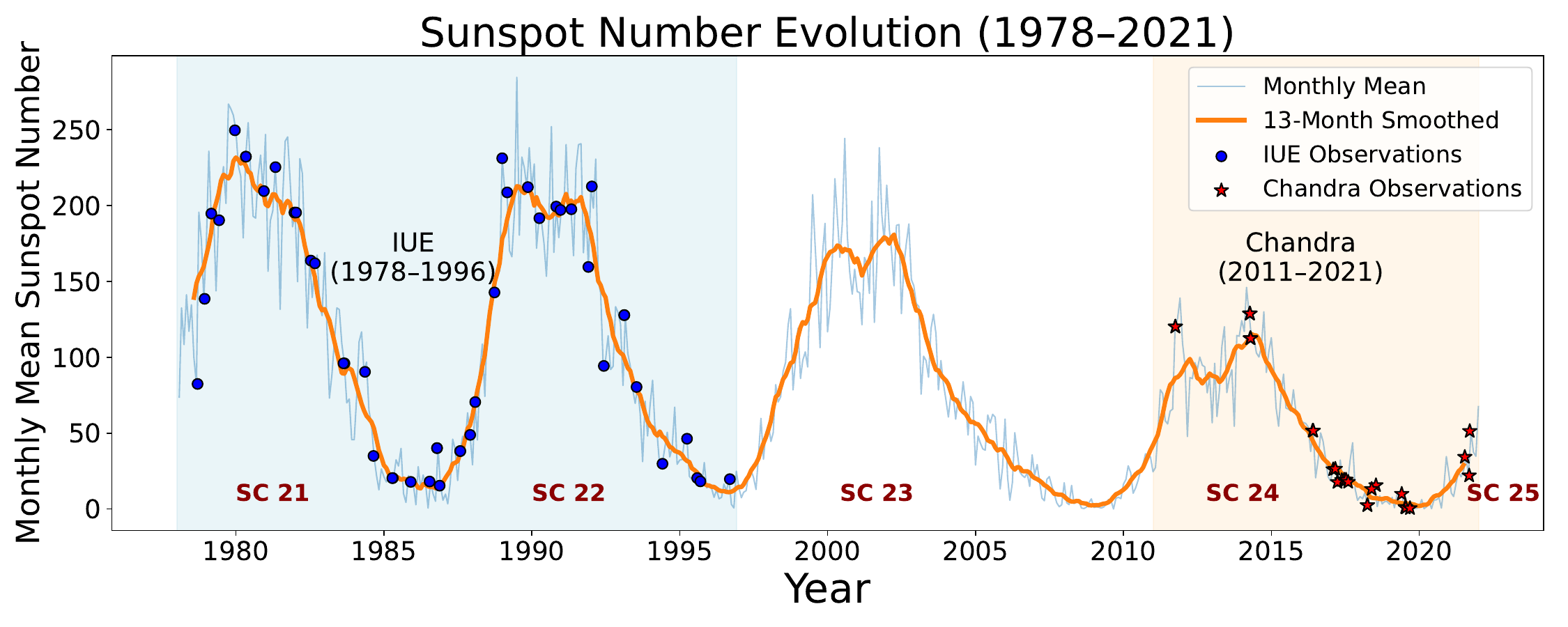}
    \caption{Evolution of the monthly averaged sunspot number from 1978 to 2021 derived from the daily total sunspot number dataset obtained from \textsc{silso}. The blue thin line represents the monthly mean sunspot number, while the orange thick curve shows the 13-month smoothed variation, highlighting the evolution of the solar cycle. Shaded regions indicate the observational periods corresponding to \IUE\ (1978–1996) and \chandra\ (2011–2021). Blue circles denote the \IUE{} observation dates, while red stars represent the \chandra{} observation epochs. The labels SC 21–SC 25 indicate Solar Cycles 21 to 25.}
    \label{fig:1}
\end{figure*}

Solar activity was characterised using contemporaneous solar X-ray flux measurements together with the sunspot number, both of which serve as widely used diagnostics of solar variability and solar-cycle evolution. Solar X-ray flux data obtained from the \GOES\ satellites\footnote{\url{https://ssd.jpl.nasa.gov/horizons/app.html##/}}, operated by the National Oceanic and Atmospheric Administration (\textit{NOAA}), were used as a proxy for coronal activity. The \GOES\ satellites are equipped with an X-ray Sensor (\textsc{xrs}) that continuously monitors the solar X-ray flux in two wavelength bands: a short channel (0.05–0.4 nm) and a long channel (0.1–0.8 nm). In this work, we used the 1-minute averaged \textsc{xrs} data obtained from the \textit{NOAA} National Centres for Environmental Information (\textit{NCEI}) archive. Monthly \textsc{csv} files were downloaded and processed, and the \textsc{b\_avg} parameter corresponding to the long channel (0.1–0.8 nm), covering photon energies of approximately 1.55–12.4 keV, was used for analysis. This energy range overlaps with the bandpass observable by the \chandra, making it directly relevant for comparison with the Jovian X-ray measurements. Depending on the requirements of the analysis, the flux values were further averaged or regrouped over hourly or daily intervals. Multiple solar-cycle variations were additionally examined using the sunspot number obtained from the World Data Centre \textsc{silso} (Sunspot Index and Long-term Solar Observations)\footnote{\url{https://www.sidc.be/SILSO/datafiles}} database, maintained by the Royal Observatory of Belgium. The \textsc{silso} dataset provides a standardised and widely used measure of solar-cycle variability. Figure \ref{fig:1} places the observational epochs of \IUE{} and \chandra{} used in this study in the context of long-term solar-cycle evolution. Our choice of data selection is driven by the unavailability of \IUE{} observations beyond 1996 and a scattered, irregular, and limited number of X-ray observations during Solar Cycle 23; hence, an observational gap may be noted in Figure \ref{fig:1}. Hence, the UV and X-ray datasets in this work are non-contemporaneous.

\begin{figure}
    \centering
       \begin{subfigure}[t]{0.48\textwidth}
        \centering
        \includegraphics[width=\textwidth]{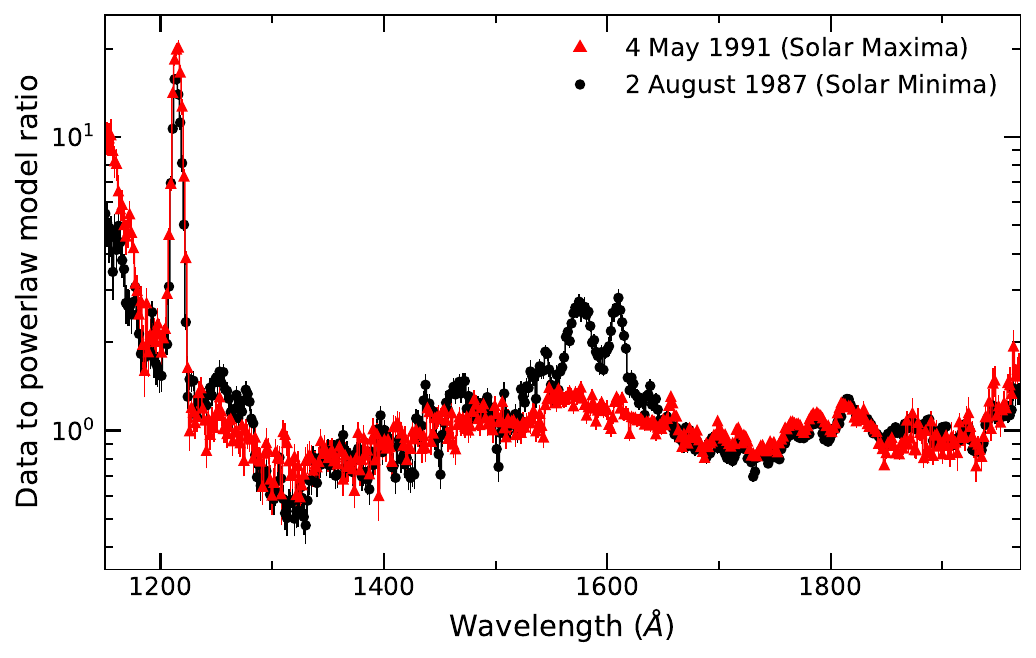}
    \end{subfigure}
    \hfill
     \begin{subfigure}[t]{0.47\textwidth}
        \centering
        \includegraphics[width=\textwidth]{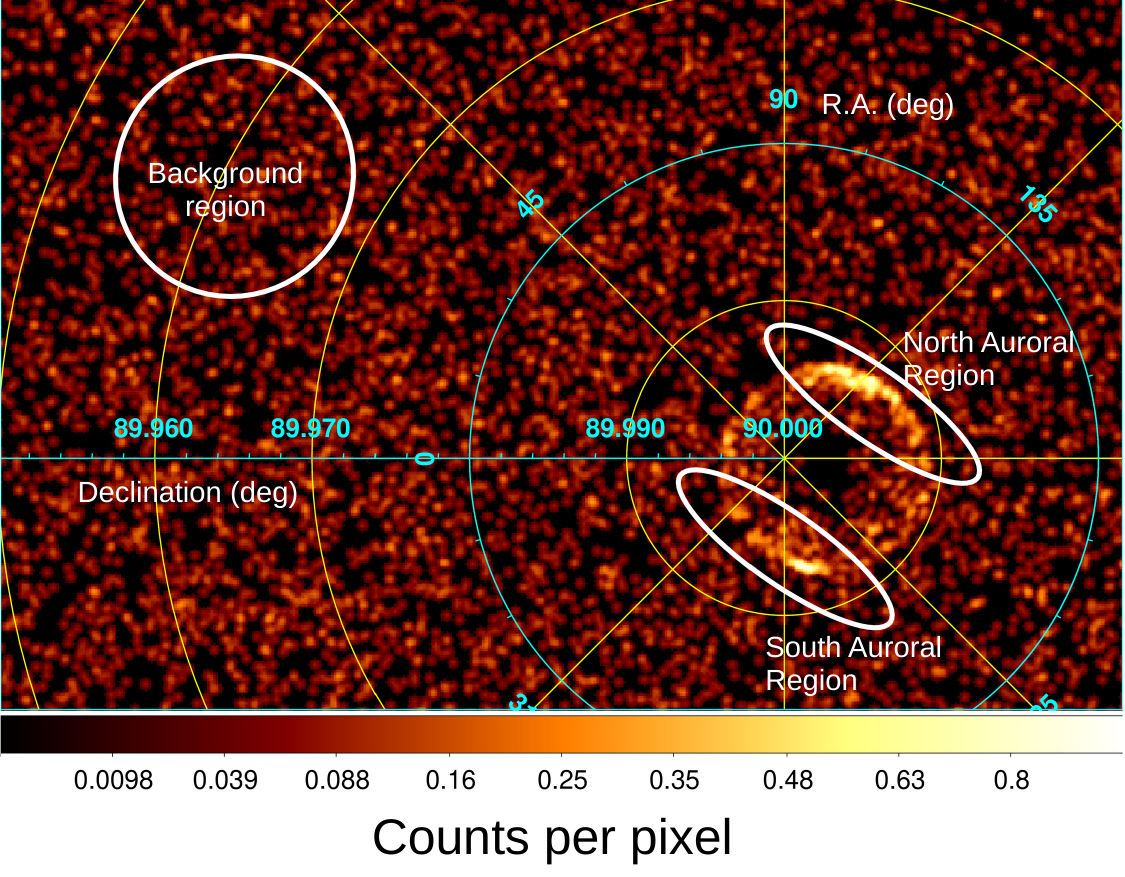}
    \end{subfigure}
    
    \caption{Top panel shows two UV ratio spectra (ratio of observed spectral counts and powerlaw model fitted counts) observed during 1987 (solar minima) and 1991 (solar maxima). Notably, double Fe\textsc{ii} line complexes near 1600~\AA~ are apparent during the solar minima. The bottom panel presents the \textit{Chandra}/ACIS image of Jupiter extracted in the 0.5-10 keV energy range. The source extraction regions corresponding to the northern and southern auroral emissions are marked, along with the background region used for background subtraction. The right ascension (RA) and declination (Dec) coordinates are also indicated on the image.}
    \label{fig:2}
\end{figure}

\begin{figure}
    \centering
       \begin{subfigure}[t]{0.45\textwidth}
        \centering
        \includegraphics[width=0.7\textwidth, angle=270]{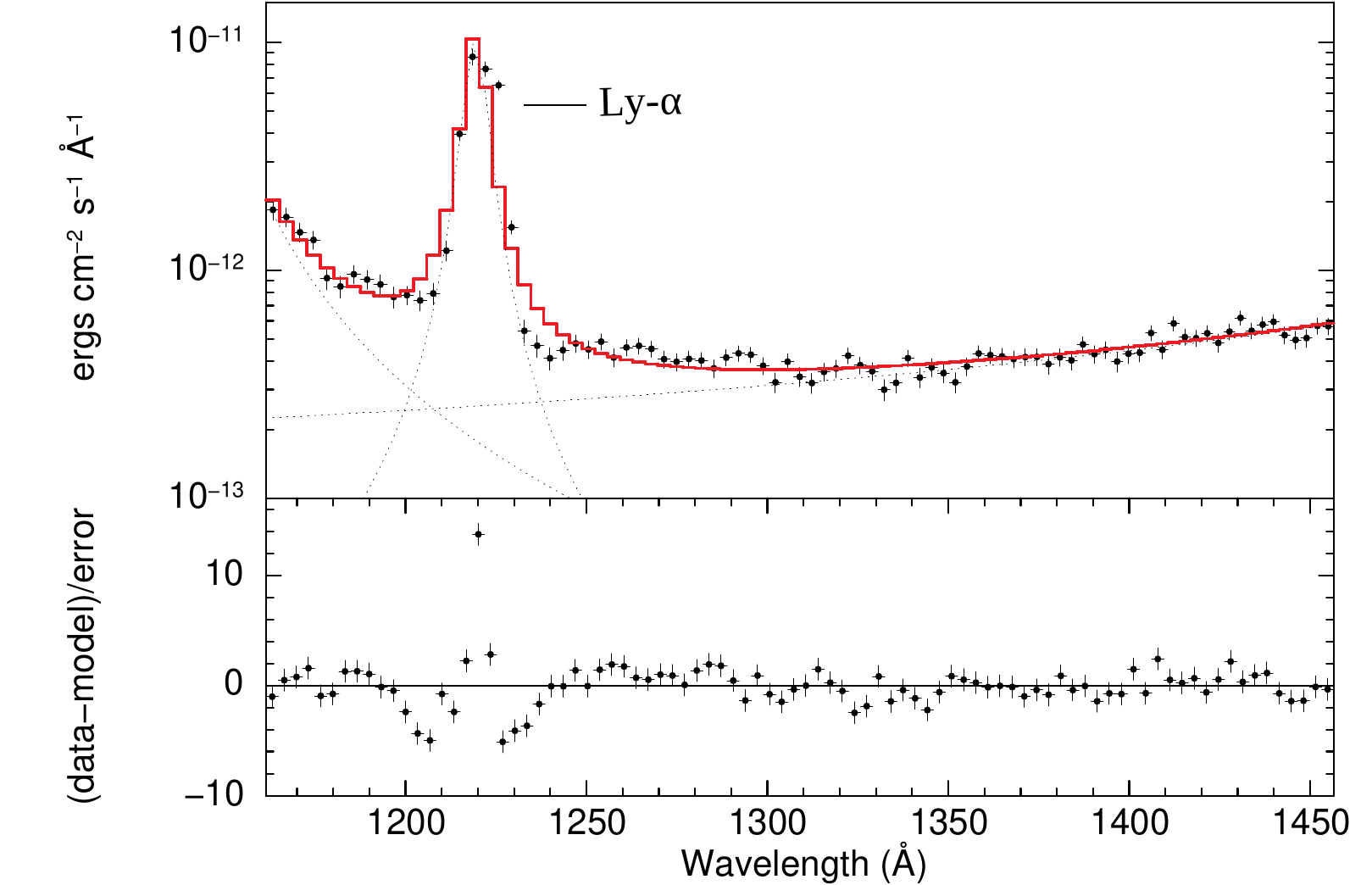}
    \end{subfigure}
    \hfill
       \begin{subfigure}[t]{0.48\textwidth}
        \centering
        \includegraphics[width=0.7\textwidth, angle=270]{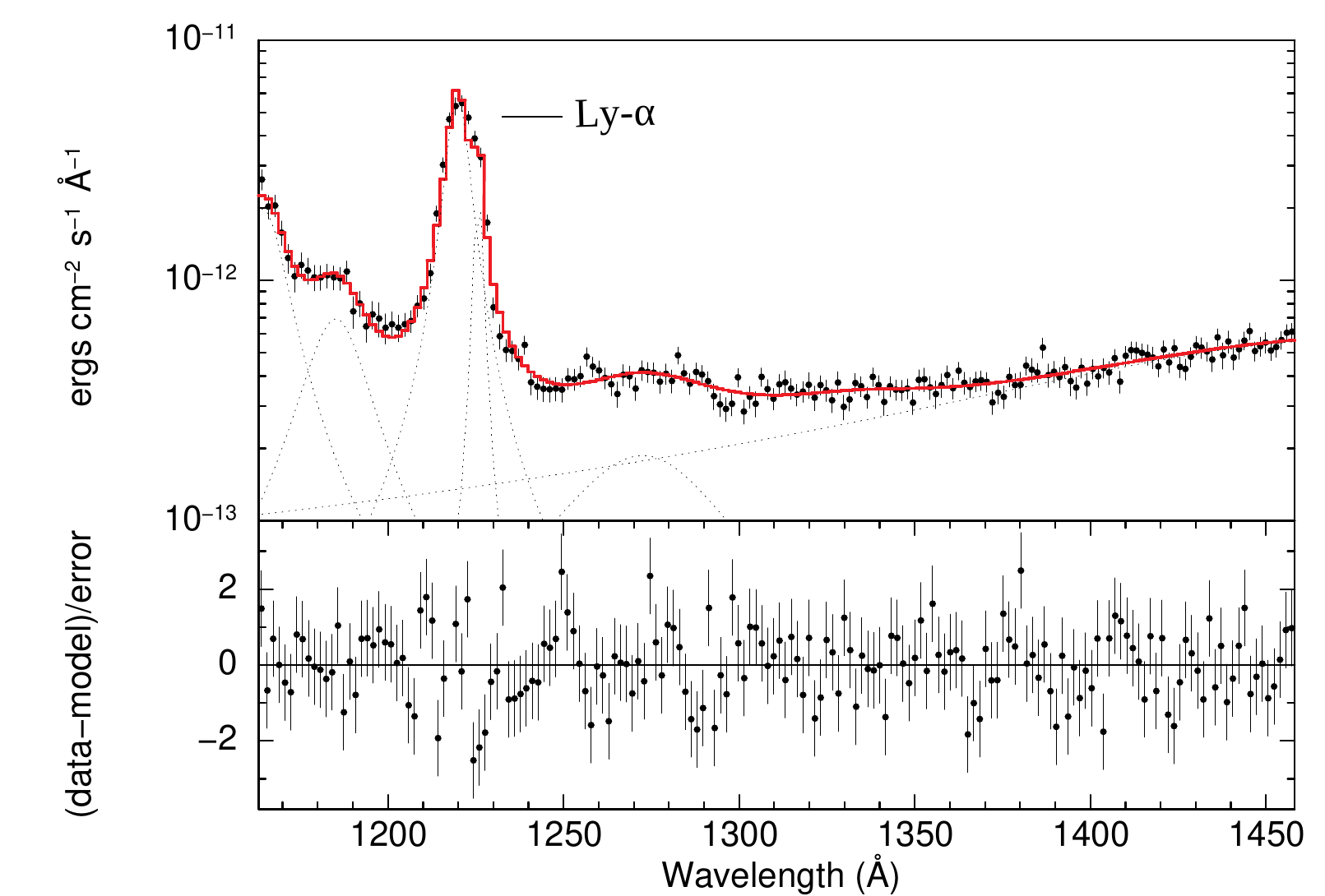}
    \end{subfigure}
    \caption{{\it A representative UV spectrum from Jupiter fitted with continuum and line models along with residuals} Top panel: An example of a poor fit ($\chi^2$/d.o.f. = 777.38/162) with a single Ly$\alpha$ line (SWP23780), illustrating the complex excess in the residuals near 1220 \AA. Bottom panel: The best-fit ($\chi^2$/d.o.f. = 164.67/155) UV spectrum (1160–1460~\AA) is shown for the same observation. The best-fit model (red) includes continuum and emission-line features, and the fitting residuals are shown in the bottom panel. The model-fitting procedure, similar to that shown in the bottom panel, was applied to all individual UV spectra.}
    \label{fig:3}
\end{figure}

\begin{table*}
\centering
\caption{\IUE\ observations of Jupiter. The top segment lists the observation IDs used for Ly$\alpha$ line analysis. The middle segment  contains the observation IDs used for Fe\textsc{ii} analysis, while the bottom segment lists the observation in which the Si\textsc{ii} line is detected.\label{tab-iue}}
\setlength{\tabcolsep}{6pt}
\small
\begin{tabular}{ccc|ccc}
\hline
Data ID & Obs Start Time & Exp Time & Data ID & Obs Start Time & Exp Time\\
 & (YYYY-MM-DD HH:MM:SS) & (s) & & (YYYY-MM-DD HH:MM:SS) & (s)\\
\hline
\hline
SWP02648	&1978-09-14 06:32:00 & 	120&    	SWP32445	&1987-12-02 21:44:47 & 	900\\	
SWP03561    &1978-12-10 00:44:04 & 	900&		SWP32839	&1988-02-02 02:16:49 & 	900\\	
SWP04465	&1979-03-03 23:39:00 & 	900&		SWP34328	&1988-09-27 01:50:21 & 	900\\	
SWP05453    &1979-06-07 20:47:17 & 	900&		SWP35215	&1988-12-31 22:36:49 & 	3900\\	
SWP07438    &1979-12-19 13:52:45 & 	135&		SWP35662	&1989-03-03 12:09:57 & 	900\\	
SWP08893	&1980-05-02 22:52:24 & 	900&		SWP37580	&1989-11-13 09:33:23 & 	900\\	
SWP08901    &1980-05-03 16:40:34 &  900&		SWP38516	&1990-04-04 18:32:36 & 	900\\	
SWP10800	&1980-12-11 13:23:07 &  540&		SWP39992	&1990-10-29 10:14:12 & 	900\\	
SWP13871	&1981-05-03 01:44:10 &  480&		SWP41564	&1991-05-04 17:25:55 & 	900\\		 
SWP15854	&1981-12-23 09:20:59 &  900&		SWP43269	&1991-11-30 07:12:11 & 	900\\	 
SWP17417	&1982-07-13 22:03:08 &  540&		SWP43613	&1992-01-12 01:38:39 & 	900\\	 
SWP17837	&1982-09-02 13:53:08 &  900&		SWP44889	&1992-06-08 16:06:00 & 	900\\	
SWP20888	&1983-09-02 13:00:57 &  720&		SWP46962	&1993-02-16 02:19:36 & 	900\\	
SWP23780	&1984-08-25 17:54:46 &  900&		SWP48136	&1993-07-17 09:52:41 & 	900\\	 
SWP27186	&1985-11-28 09:32:38 &  900&		SWP50984	&1994-06-02 14:20:27 & 	900\\	
SWP28712	&1986-07-19 14:55:45 &  900&		SWP54274    &1995-04-01 21:44:32 &  900\\		 
SWP29473	&1986-10-18 07:02:37 &  900&		SWP55426    &1995-08-05 07:45:21 &  1800\\
SWP31439	&1987-07-31 17:20:59 & 	900&        SWP55919    &1995-09-13 02:34:53 & 1800\\
SWP31449	&1987-08-02 08:02:51 & 	900&        SWP58210    &1996-09-11 10:43:17 & 1800\\
\hline
\hline
SWP16025	&1982-01-12 04:17:58 & 2700& SWP29723	&1986-11-20 04:30:05 &  900\\
SWP20739	&1983-08-21 14:35:02 &  900& SWP31444	&1987-08-01 12:32:02 &  900\\
SWP23004	&1984-05-12 21:25:06 &  900& SWP37074	&1989-09-19 03:39:41 & 1800\\
SWP22998	&1984-05-12 16:31:45 &  900& SWP40417	&1990-12-21 02:12:52 &  900\\
SWP25754	&1985-04-21 00:24:17 &  900& SWP46952	&1993-02-15 18:02:59 & 1200\\
SWP29482	&1986-10-19 07:45:25 & 1800& SWP48130	&1993-07-17 05:34:20 &  900\\
\hline
\hline
SWP25688	&1985-04-15 02:27:56 & 600	&	& &  \\
\hline
\end{tabular}
\end{table*}

\subsection{\IUE\ observation and data reductions}
The \IUE\ obtained a substantial number of low-resolution short-wavelength spectra of Jupiter between September 13, 1978, and September 25, 1996, covering the 1150–1978~\AA~wavelength range. For data reduction and analysis, we follow the methodology of \citet{1997ApJS..113...69W}. The raw spectral images are processed using both the \textsc{tomsips} \citep{1993PASP..105..538A} and \textsc{newsips} \citep{1996AJ....111..517N} pipelines. In line with the recommendation of \citet{1997ApJS..113...69W}, we adopt the \textsc{newsips} reductions, since the \textsc{tomsips} spectra suffer from wavelength calibration nonlinearities associated with uncorrected long-term drifts. By contrast, the \textsc{newsips} products are in agreement with \HST{} observations \citep{1997ApJS..113...69W}. However, a 10\% uncertainty has been incorporated in the line and continuum flux measurements from spectral modelling as pointed out in earlier analysis with \textsc{newsips} reductions \citep{2000ApJS..126..517M}. Each spectrum is examined in the 1160–1460~\AA~window to exclude cases with high noise levels, saturation, or inadequate exposure. After applying these quality checks and taking into account the availability of concurrent solar data, 51 spectra are selected for subsequent analysis, which are listed in Table \ref{tab-iue}. Example UV spectra during solar maxima and solar minima are shown in the top panel of Figure \ref{fig:2}. The observation dates chosen in Figure \ref{fig:2} are mere representations of all 51 UV spectra to show how UV emission from Jupiter behaves in terms of spectral shape and line properties during Solar minima and maxima. For clarity of short-term continuum variability and better visibility of line features, both spectra are fitted with a single powerlaw and the spectral counts to powerlaw fitted model counts ratio is shown as a function of wavelength. While details of spectral analysis have been provided later, a substantial line and continuum variation can be observed in the UV regime between solar maxima and minima.

\subsection{\textit{Chandra}}
Jupiter has been observed extensively with \chandra\ using the Advanced CCD Imaging Spectrometer (ACIS) and the High Resolution Camera (HRC). The \textit{Chandra}/ACIS observations were carried out in ACIS-S mode using the back-illuminated ACIS-S3 chip having 1024 $\times$ 1024  pixels with a pixel size of approximately 0.492 arcseconds, allowing spatially resolved spectroscopy of Jupiter's auroral regions over the photon energy range of 0.2–10 keV \citep{2003SPIE.4851...28G}. In this work, we have focused on the 2011 observations of Jupiter through \textit{Chandra}/ACIS, which are summarised in Table~\ref{Tab1}.

{
\begin{table*}
\captionsetup{justification=centering, singlelinecheck=false}
\centering
\caption[]{ \textit{Chandra}/ACIS Observations of Jupiter, where NP and SP are the north pole and south pole, respectively. The revised count rates are obtained after distance ($d_{\rm EJ}$) normalisation }\label{Tab1}

\begin{tabular}{cccccccc}
  \hline\noalign{\smallskip}

No &  Obs ID  & Obs Date & Start time & Exposure time  & $d_{\rm EJ}$ &  Revised NP   & Revised SP  \\
& & & &  (ksec) & (AU) &  count rate(x10$^{-3}$)&  count rate(x10$^{-3}$)  \\
\hline\noalign{\smallskip}
obs1 & 12315 & 2011-10-02 & 21:54:26 & 39.25 & 4.06 & 10.17$\pm{0.66}$ & 2.23$\pm{0.64}$ \\
obs2 & 12316 & 2011-10-04 & 14:33:08 & 39.61 & 4.05 & 5.99$\pm{0.54}$ & 1.78$\pm{0.49}$ \\
  \noalign{\smallskip}\hline
\end{tabular}
\end{table*}
}
The \textit{Chandra}/HRC observations were carried out in HRC-I mode, between 2014 and 2021, which captured auroral emission, morphology, and variability. Details of observations are provided in Table~\ref{Tab2}, forming a rich dataset for studying Jupiter’s X-ray auroral activity.

\begin{table*}
\centering
\caption{\textit{Chandra}/HRC Observations of Jupiter with distance-normalised count rates scaled to the first observation’s Jupiter distance}\label{Tab2}
\footnotesize
\begin{tabular}{cccccccc}
\hline
No & ObsID & Obs Date & Start Time & Exposure  & Average Count Rate  & $d_{\rm EJ}$  & Revised Count Rate \\
&&&& (ksec) & ($10^{-3}$ s$^{-1}$) & (AU) & ($10^{-3}$ s$^{-1}$) \\
\hline
\hline
obs1  & 15671 & 2014-04-08 & 08:18:10 & 40.73 & $25 \pm 2$ & 5.24 & $25.00 \pm 2.00$ \\
obs2  & 16299 & 2014-04-10 & 01:09:23 & 37.53 & $15 \pm 1$ & 5.27 & $14.85 \pm 0.99$ \\
obs3  & 15672 & 2014-04-12 & 22:09:30 & 39.51 & $15 \pm 2$ & 5.32 & $14.59 \pm 1.95$ \\
obs4  & 15669 & 2014-04-15 & 20:43:04 & 37.85 & $11 \pm 1$ & 5.36 & $10.54 \pm 0.96$ \\
obs5  & 16300 & 2014-04-17 & 12:19:31 & 39.46 & $11 \pm 2$ & 5.39 & $10.45 \pm 1.90$ \\
obs6  & 15670 & 2014-04-20 & 02:19:30 & 23.12 & $15 \pm 3$ & 5.43 & $13.98 \pm 2.80$ \\

obs7 & 18608 & 2016-05-24 & 10:21:59 & 39.74 & $11 \pm 2$ & 5.17 & $11.30 \pm 2.05$ \\
obs8 & 18609 & 2016-06-01 & 11:31:01 & 39.72 & $9 \pm 2$ & 5.30 & $8.82 \pm 1.96$ \\

obs9  & 18301 & 2017-02-02 & 09:56:57 & 32.73 & $9 \pm 2$ & 5.03 & $9.77 \pm 2.17$ \\
obs10 & 18302 & 2017-05-19 & 00:27:33 & 37.32 & $14 \pm 2$ & 4.69 & $18.27 \pm 2.61$ \\

obs11 & 20000 & 2017-02-28 & 12:38:55 & 70.60 & $7 \pm 2$ & 4.68 & $8.78 \pm 2.51$ \\
obs12 & 18676 & 2017-03-27 & 08:30:56 & 9.30 & $9 \pm 3$ & 4.48 & $12.31 \pm 4.10$ \\
obs13 & 20001 & 2017-06-18 & 18:37:56 & 36.67 & $8 \pm 2$ & 5.09 & $8.48 \pm 2.12$ \\
obs14 & 18677 & 2017-07-10 & 21:11:18 & 39.56 & $9 \pm 1$ & 5.43 & $8.34 \pm 0.93$ \\
obs15 & 20002 & 2017-08-06 & 01:55:48 & 34.98 & $12 \pm 3$ & 5.81 & $9.73 \pm 2.43$ \\

obs16  & 18303 & 2018-07-15 & 23:43:48 & 40.03 & $12 \pm 1$ & 4.96 & $13.42 \pm 1.12$ \\
obs17 & 18678 & 2018-03-31 & 23:10:42 & 40.05 & $9 \pm 2$ & 4.63 & $11.55 \pm 2.57$ \\
obs18 & 20733 & 2018-04-01 & 10:38:00 & 38.79 & $15 \pm 3$ & 4.62 & $19.29 \pm 3.86$ \\
obs19 & 18679 & 2018-05-23 & 23:59:44 & 39.54 & $7 \pm 2$ & 4.43 & $9.80 \pm 2.80$ \\

obs20 & 22159 & 2019-05-29 & 03:26:26 & 35.23 & $9 \pm 2$ & 4.31 & $13.28 \pm 2.95$ \\
obs21 & 22146 & 2019-07-13 & 01:26:23 & 24.46 & $16 \pm 3$ & 4.42 & $22.53 \pm 4.22$ \\
obs22 & 22147 & 2019-07-13 & 21:08:30 & 22.66 & $14 \pm 3$ & 4.43 & $19.59 \pm 4.20$ \\
obs23 & 22148 & 2019-07-15 & 12:58:56 & 24.73 & $6 \pm 4$ & 4.44 & $8.37 \pm 5.58$ \\
obs24 & 22149 & 2019-07-16 & 08:44:51 & 24.80 & $18 \pm 3$ & 4.45 & $25.02 \pm 4.17$ \\
obs25 & 22150 & 2019-07-18 & 20:17:53 & 24.75 & $18 \pm 3$ & 4.47 & $24.69 \pm 4.11$ \\
obs26 & 22151 & 2019-09-08 & 22:58:16 & 24.33 & $11 \pm 2$ & 5.18 & $11.29 \pm 2.05$ \\

obs27 & 23370 & 2021-07-16 & 14:02:24 & 35.69 & $20 \pm 4$ & 4.18 & $31.40 \pm 6.28$ \\
obs28 & 23366 & 2021-09-08 & 11:17:15 & 35.87 & $55 \pm 5$ & 4.07 & $91.46 \pm 8.31$ \\
obs29 & 23369 & 2021-09-15 & 12:32:16 & 140.19 & $67 \pm 5$ & 4.11 & $108.67 \pm 8.11$ \\

\hline
\end{tabular}
\end{table*}

\subsubsection{ACIS data reduction}
Data reduction was performed with the Chandra Interactive Analysis of Observations (\textsc{ciao}) software \textsc{version 4.17.0} using Calibration Database (\textsc{caldb}) \textsc{version 4.12.0} \citep{2006SPIE.6270E..1VF}. The $\textsc{chandra\_repro}$ pipeline was executed for each ObsID with the \textsc{sso\_freeze} option enabled to correct for any apparent motion of Jupiter during the observation. The X-ray image of Jupiter in the bottom panel of Figure \ref{fig:2} clearly indicates that the northern auroral region shows the brightest X-ray emission. 

For both the northern and southern poles, elliptical source region files were created in \textsc{ds9} with \textsc{[bin=ocx,ocy]}, while background regions of a similar shape and size were selected from areas above the poles (to exclude contributions from Io's plasma torus), which are away by three times the diameter of Jupiter's disc. For consistency throughout the \chandra{} analyses, elliptical source regions of similar sizes (semi-major axis of 30$''$  and semi-minor axis of 8$''$) were adopted for both the northern and southern auroral emissions. The coordinates and dimensions of these elliptical regions were selected such that they consistently encompass the auroral zones extending from the poles down to approximately $75^\circ$ latitude in both hemispheres. This choice ensures that the majority of the X-ray counts originating from the polar auroral regions are included, thereby providing robust measurements of the count rates and X-ray spectral properties. The large angular separations between the background region centre and the centres of the northern and southern auroral ellipses (165$''$ and $155''$, respectively) are adopted to make sure background region counts are not contaminated by source counts. The source and background regions together with coordinate systems are illustrated in the bottom panel of Figure \ref{fig:2}.
To identify the locations of Jupiter's northern and southern poles in the \chandra{} images, ephemeris data were generated using the NASA JPL Horizons System\footnote{\url{https://ssd.jpl.nasa.gov/horizons/app.html##/}} corresponding to each observation. Using parameters like the apparent right ascension and declination of the centre of Jupiter's disc together with the sub-observer and sub-solar longitudes and latitudes, as well as parameters such as the north-pole position angle in degrees, angular diameter of disc in arcsec, and sky motion, we determine the apparent orientation and viewing geometry of Jupiter during each observation, and identify both poles in \chandra{} images. 
Spectral extraction was performed with the \textsc{specextract} task, which generated the associated auxiliary response files (ARFs), redistribution matrix files (RMFs), along with background-subtracted pulse invariant (PI) files. For each spectrum, spectral channels are binned such that the minimum signal-to-noise ratio of each spectral bin is 3. With this criterion, each spectrum contains a minimum of 20 spectral bins for further modelling.

Backscale corrections were applied to both source and background spectra based on the areas of the defined regions. All the net count rates derived for the northern and southern auroral regions from these spectral extractions are summarised in Table~\ref{Tab1}. For the 2011 \textit{Chandra}/ACIS observations, in addition to spectral extraction, a background-subtracted light curve was also generated for the northern auroral region using the \textsc{ciao} tool \textsc{dmextract}. The source light curve was extracted from the northern auroral region with a time bin size of 1024 seconds. With the 1024 sec bin size, we obtain a minimum signal-to-noise ratio of 2 for each count rate bin. The extracted light curve was then corrected for background and used for further analysis.

\subsubsection{HRC data reduction}
Similar to the \textit{Chandra}/ACIS data, the initial steps of reduction for \textit{Chandra}/HRC observations were carried out with the \textsc{ciao} software using \textsc{caldb}, and for each observation, the \textsc{chandra\_repro} pipeline was executed to produce Level 2 event files. These event files were subsequently examined in \textsc{ds9} to define the auroral source and background extraction regions. Throughout the \chandra{}/HRC analyses, elliptical source regions of similar sizes (semi-major axis of 9$''$  and semi-minor axis of 3.5$''$) were adopted for both the northern and southern auroral emissions, while the circular background region of radius 8.3$''$ is chosen at a separation of 55$''$ as shown in the top left panel of Figure \ref{fig:7}.

The southern auroral region consistently showed poor photon statistics across all 29 \textit{Chandra}/HRC observations; no source is detected above the background count rate of 1.43$\times$10$^{-3}$ counts/sec (determined from a source-free, circular region which is sufficiently away from Jupiter's disc) with at least 2$\sigma$ significance. Therefore, only the northern auroral emission was considered for further analysis.
Light curves corresponding to the source and background regions were extracted for each dataset using the \textsc{dmextract} tool, binned at 1024-second intervals. The background-subtracted light curves were produced using \textsc{lcmath}, where the background counts were scaled and subtracted from the source counts to isolate Jupiter’s auroral emission. Light curves were then analysed using \textsc{lcstats}, from which the average count rates were calculated for each observation and shown in Table~\ref{Tab2}.

\section{Data analysis and Results}
\label{sect:data}
\begin{figure*}
\centering

\begin{subfigure}[t]{0.42\textwidth}  
    \centering
    \includegraphics[width=1.0\linewidth,height=0.27\textheight,keepaspectratio]{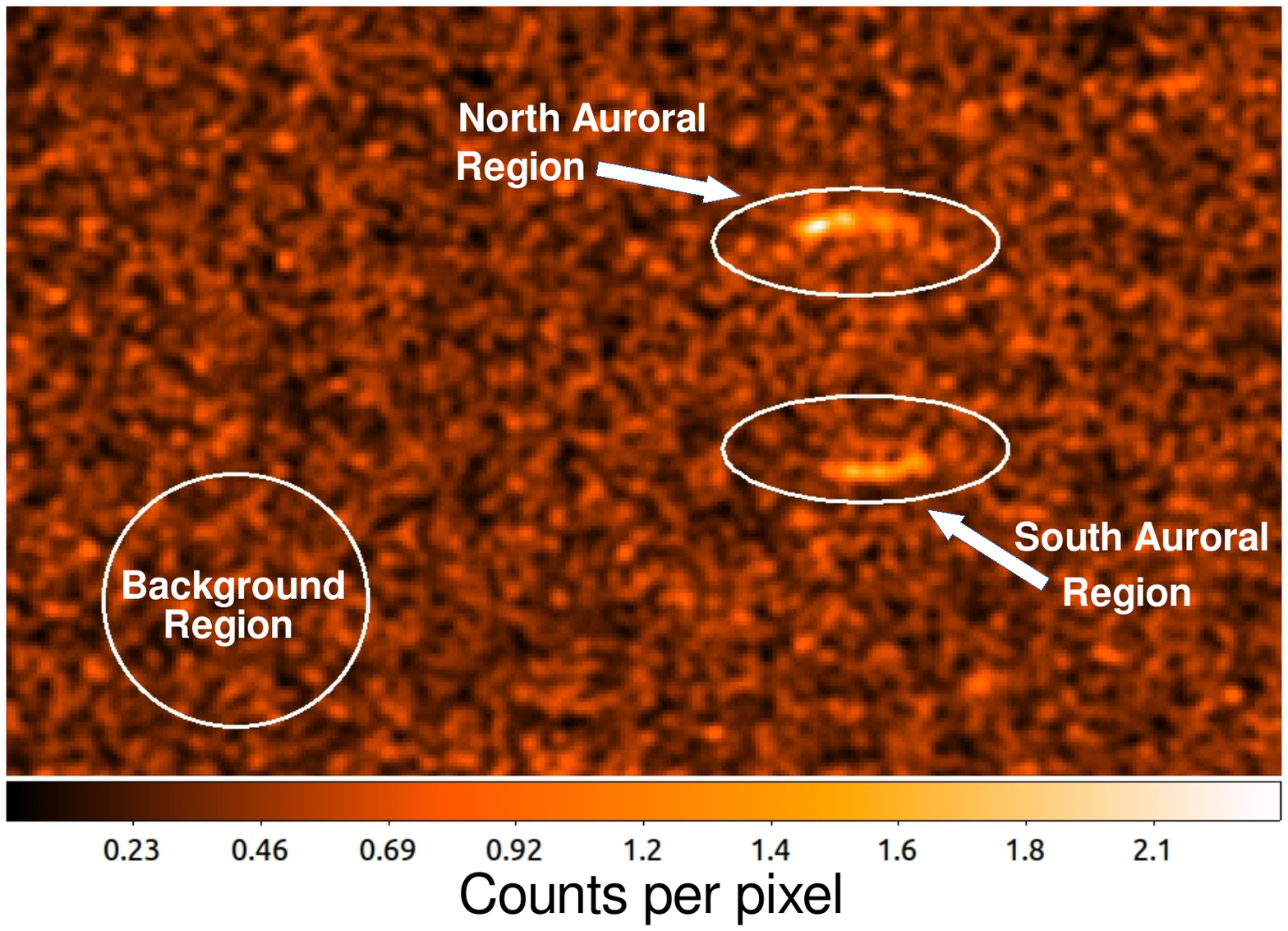}
\end{subfigure}
\hspace{0.05\textwidth}              
\begin{subfigure}[t]{0.50\textwidth}
    \centering
    \includegraphics[width=1.0\linewidth,height=0.32\textheight,keepaspectratio]{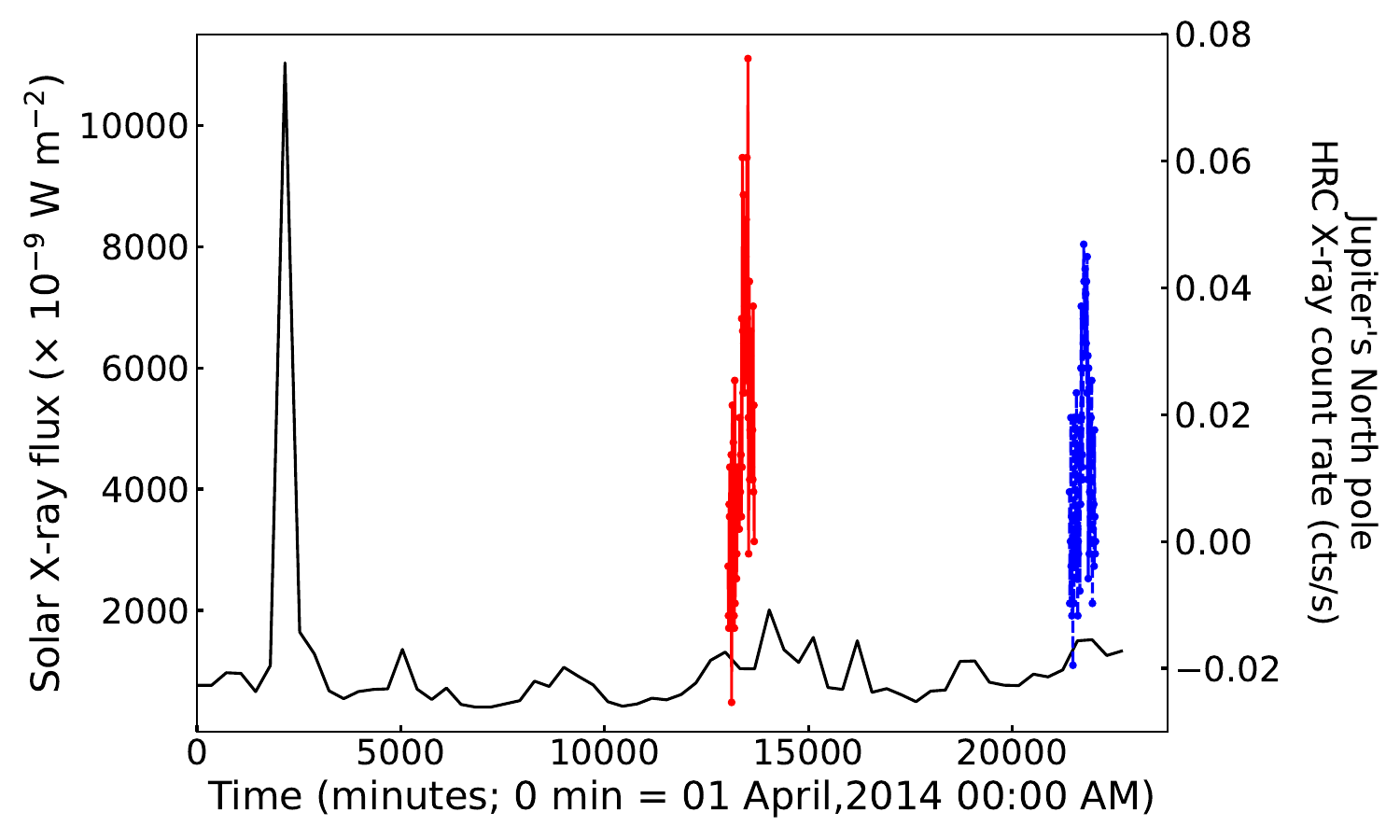}
\end{subfigure}

\vspace{0.02\textheight}              
\begin{subfigure}[t]{0.48\textwidth}
    \centering
    \includegraphics[width=1.1\linewidth,height=0.35\textheight,keepaspectratio]{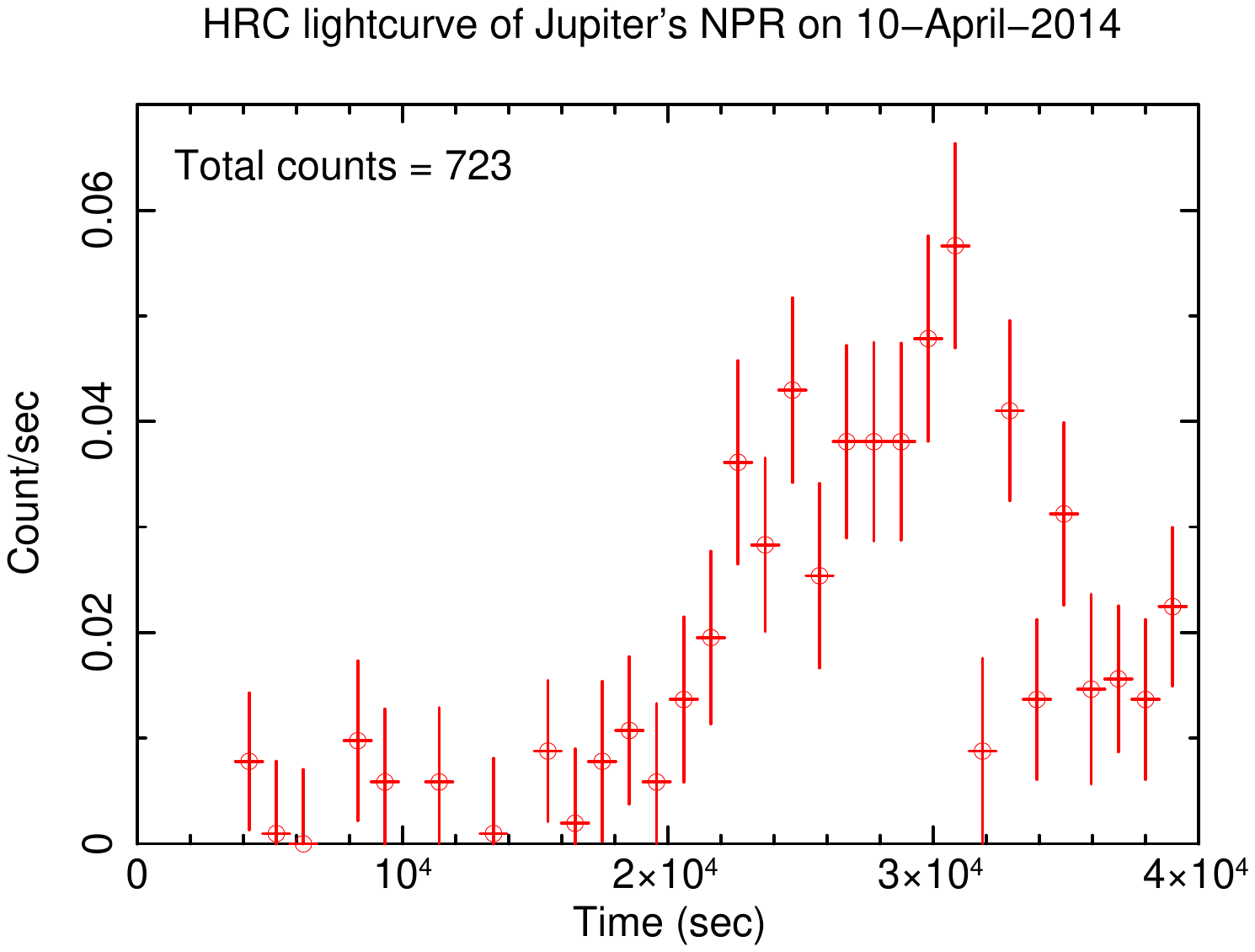}
\end{subfigure}
\hspace{0.03\textwidth}
\begin{subfigure}[t]{0.48\textwidth}
    \centering
    \includegraphics[width=1.1\linewidth,height=0.35\textheight,keepaspectratio]{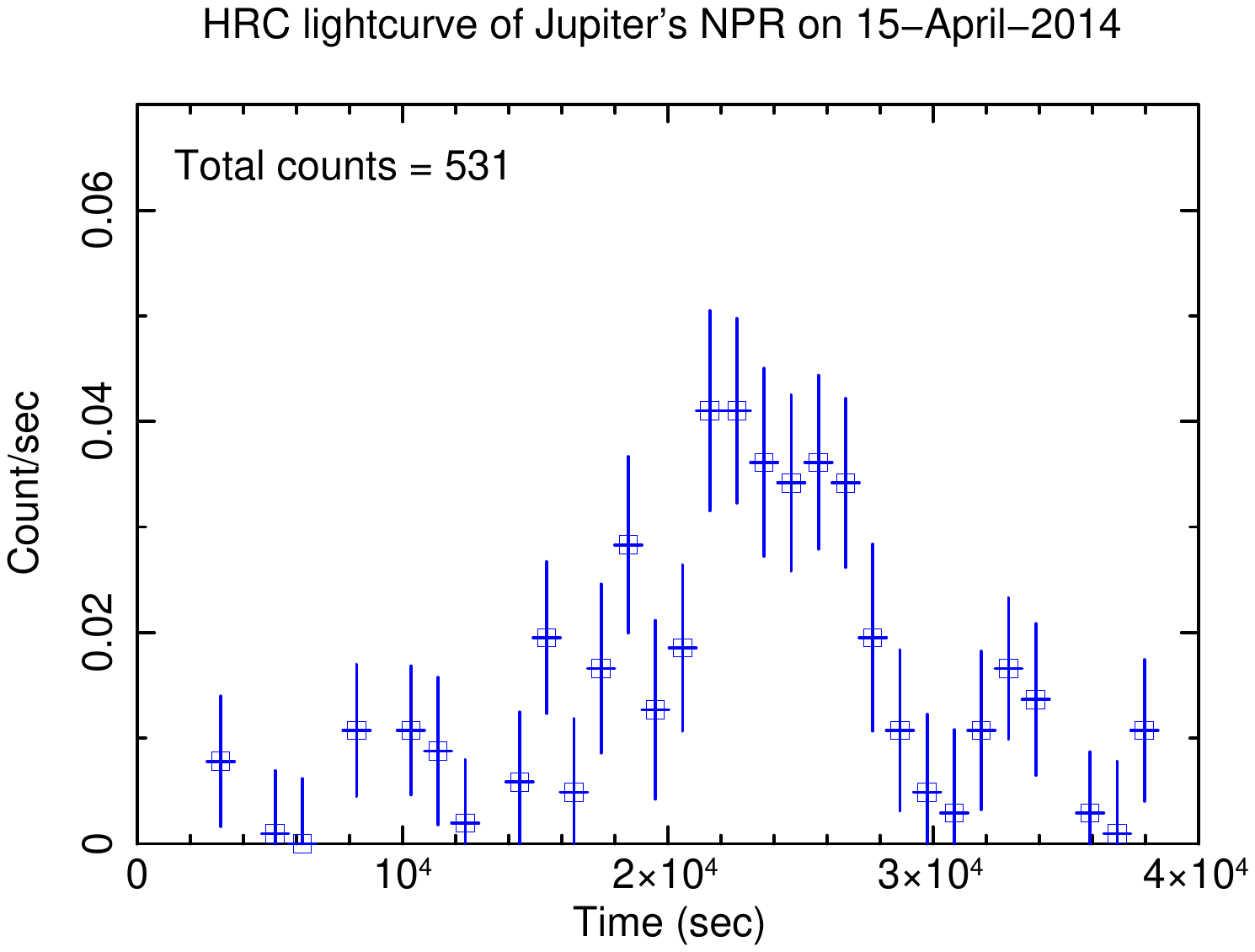}
\end{subfigure}

\caption{\textit{\chandra{}/HRC X-ray flares and CME.} The top left panel shows the \textit{Chandra}/HRC image of Jupiter in 0.5-10 keV as observed on 20th April, 2014. The source and background regions used for analysis are marked in the image. The top right panel shows the 0.5-10 keV background-subtracted \chandra{}/HRC lightcurves for Jupiter's northern auroral region as observed on 10th April, 2014 (shown in red circles) and 15th April, 2014 (shown in blue squares), along with the corresponding solar X-ray flux (in black solid line) starting from 1st April, 2014. A CME was reported and shown as observed on 03 April, 2014. For the clear visibility of X-ray flares, zoomed versions of both light curves are shown in the bottom left and bottom right panels, respectively.}
\label{fig:7}
\end{figure*}

 To illustrate the energy-dependent morphology of Jupiter’s X-ray emission, we present a \textit{Chandra}/HRC image of Jupiter (top left panel of Figure \ref{fig:7}), in which the northern and southern auroral zones have been explicitly marked to indicate the regions selected for subsequent analysis. This image was visualised in \textsc{ds9} using object-centred coordinates, ensuring that the emission is displayed relative to the Jupiter disc.

\begin{figure*}
\centering
\includegraphics[scale=0.28]{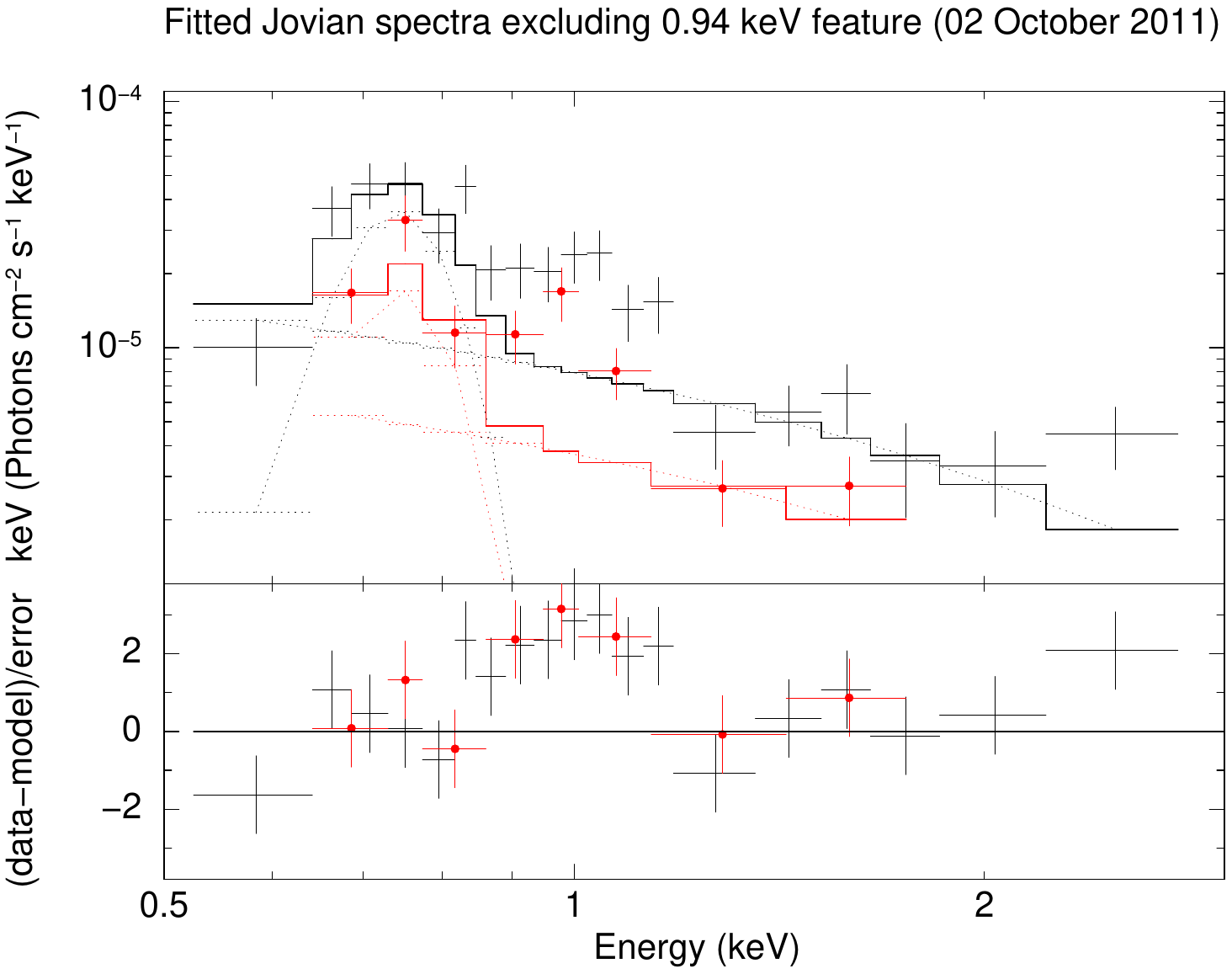}
\includegraphics[scale=0.28]{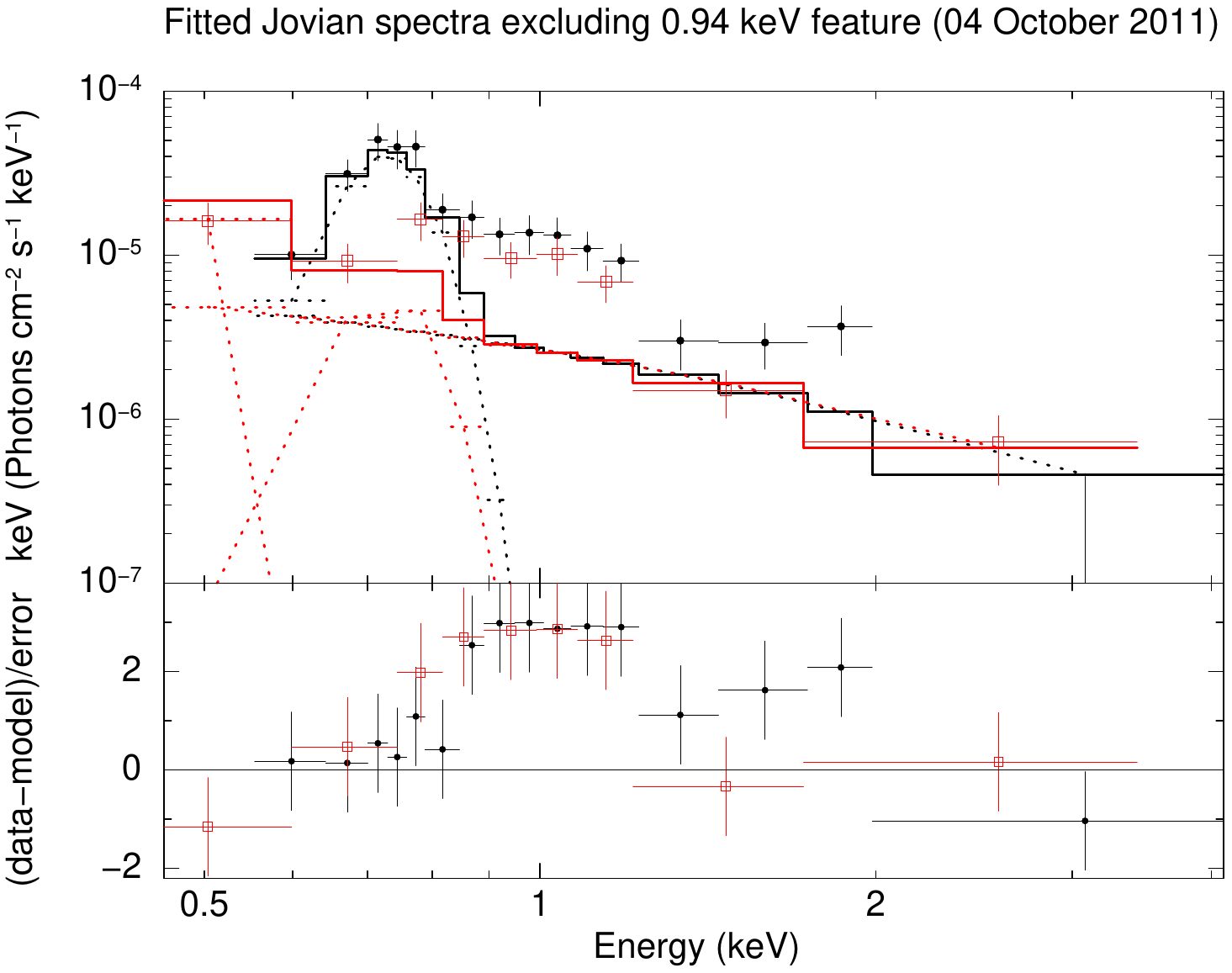}
\includegraphics[scale=0.28]{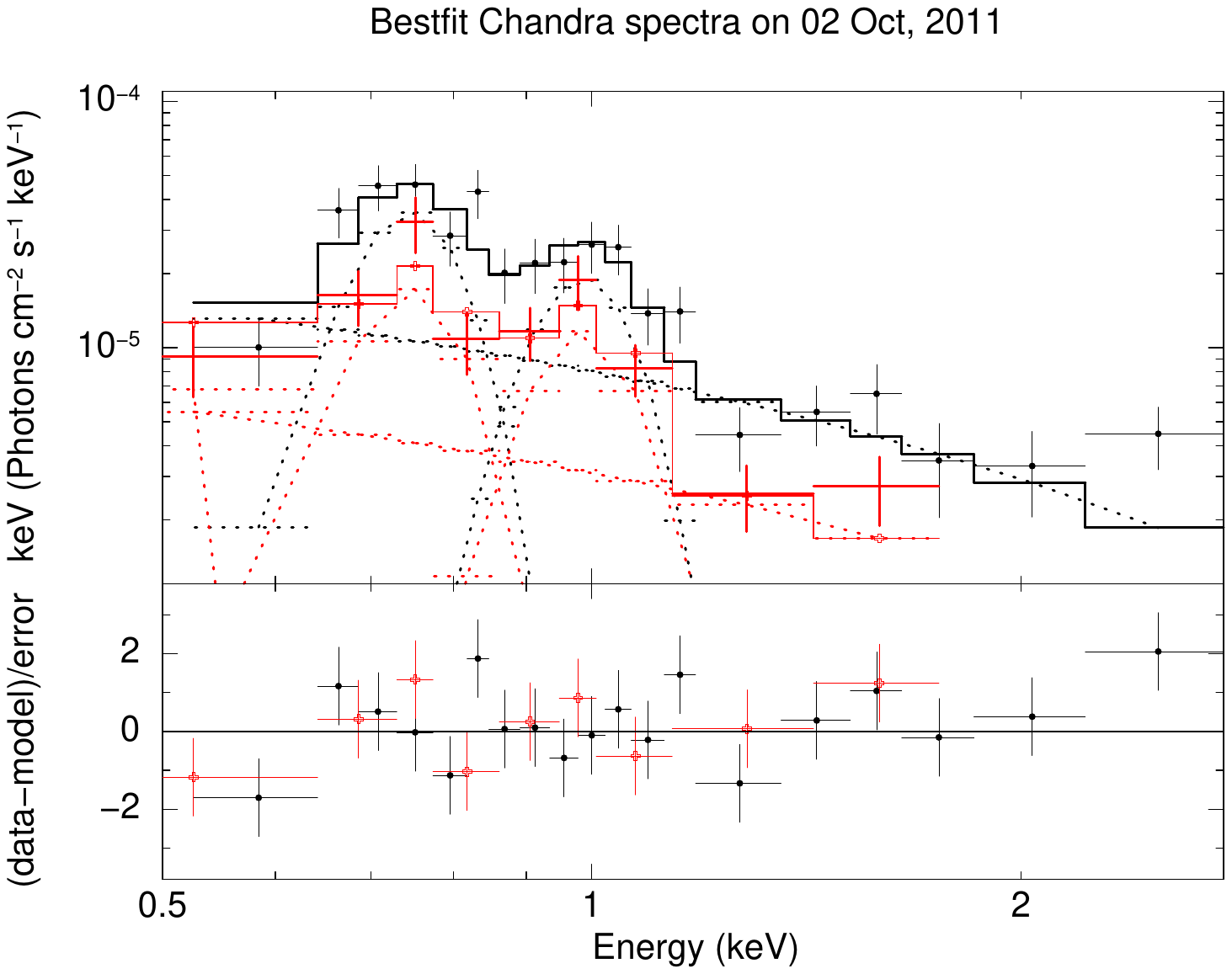}
\includegraphics[scale=0.28]{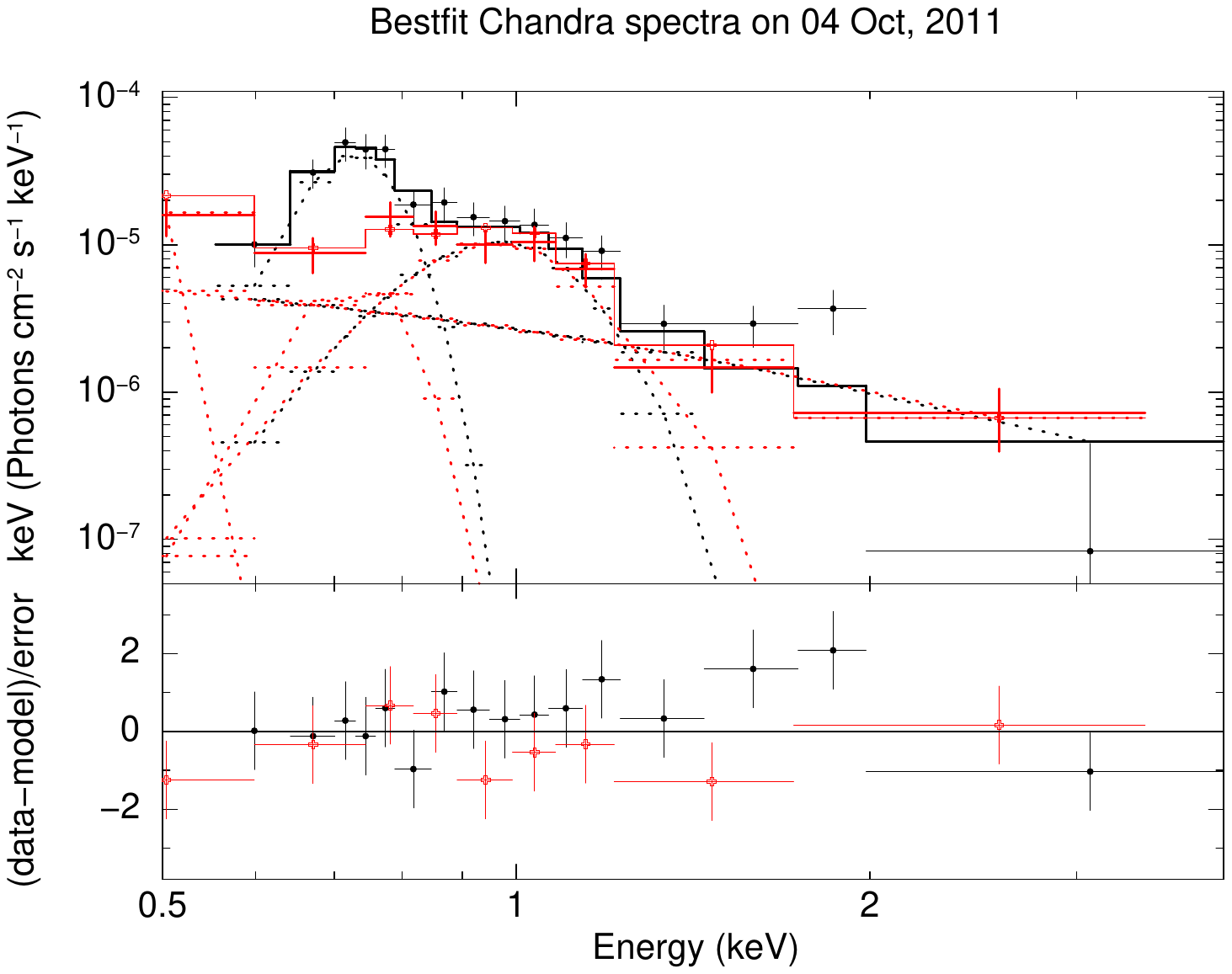}
  \caption{Top panels show \chandra{}/ACIS fitted X-ray spectra obtained from the northern polar region (black crosses) and southern polar region (red circles) of Jupiter on 2nd October 2011 (top left panel) and 4th October 2011 (top right panel), with a combination of a Bremsstrahlung and Gaussian emission line near 0.73 keV. A significant excess is observed near one keV in residual plots during both dates. The excess can be mitigated and the best-fit obtained by adding another Gaussian line near 0.94 keV, as shown in the bottom-left and bottom-right panels for 2nd and 4th October 2011, respectively. Model components used to fit all spectra are shown by thin dotted lines.}  
  \label{fig:4}
\end{figure*}

\begin{figure}
    \centering
  \begin{subfigure}{0.50\textwidth}
    \includegraphics[width=\textwidth]{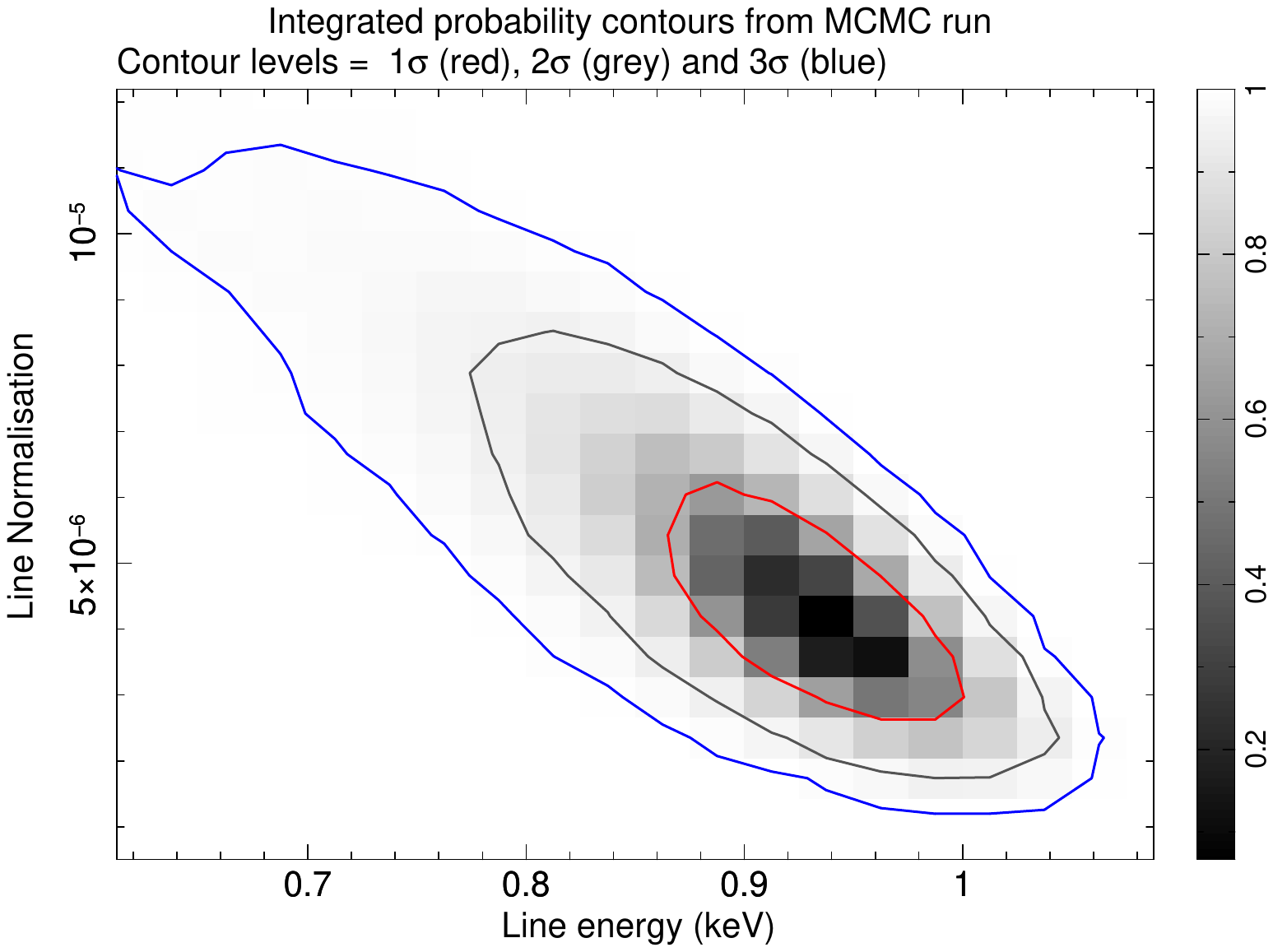}  
    \label{fig:sub1}
  \end{subfigure}%
  \hfill
  \begin{subfigure}{0.49\textwidth}
    \includegraphics[width=\textwidth]{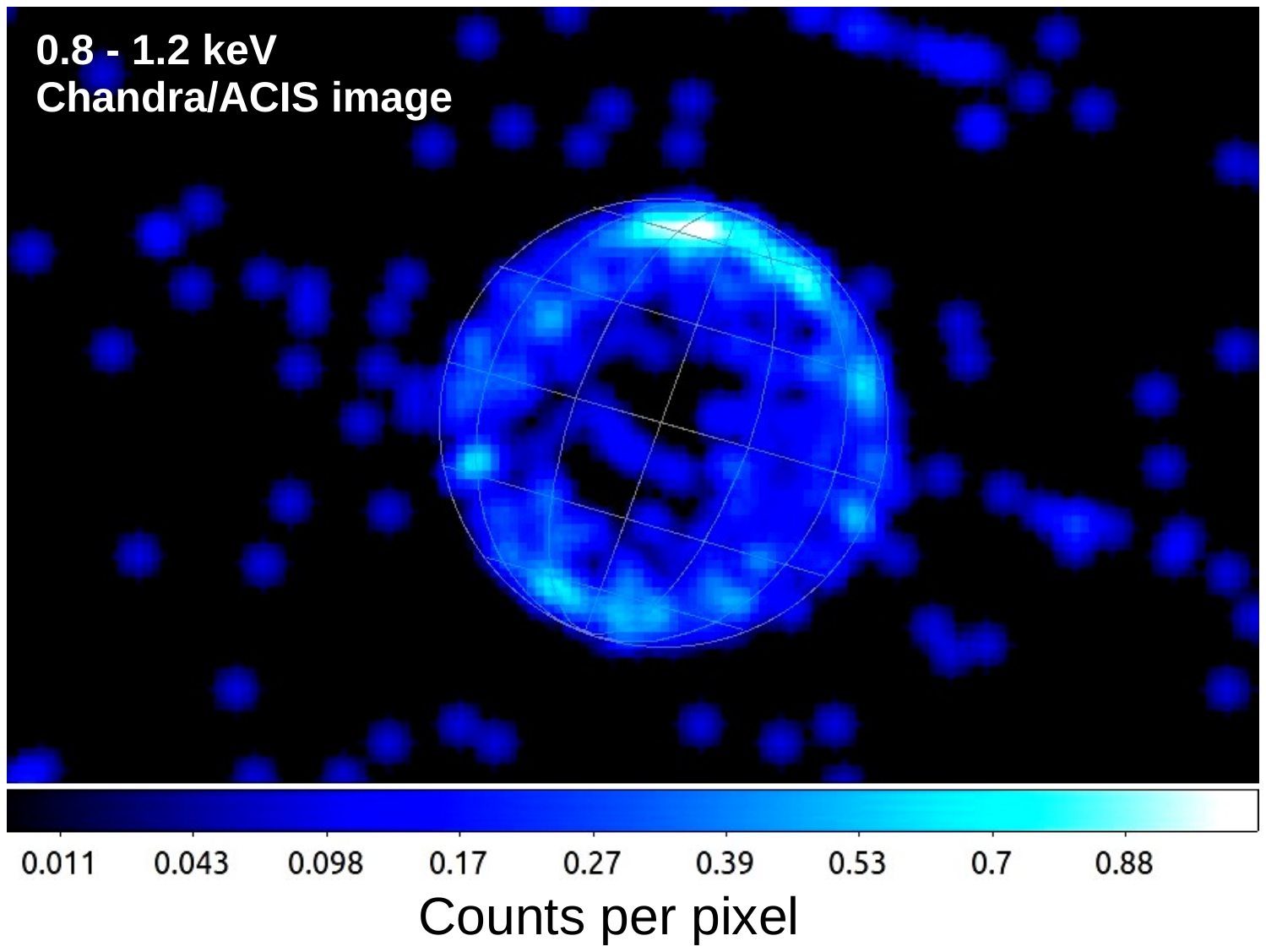}
    \label{fig:sub2}
  \end{subfigure}
  \caption{Top panel shows 1, 2 and 3$\sigma$ integrated contours from the marginal probability distribution of Ne$^{8+}$ emission line energy near 0.94 keV versus line normalisation as obtained from the best-fit X-ray spectral parameters shown in the bottom left panel of Figure \ref{fig:4}. Bottom panel shows \textit{Chandra}/ACIS image extracted in the 0.8–1.2~keV energy range from the observations conducted on 2nd October, 2011 (ObsID 12315).}
  \label{fig:5}
\end{figure}

\begin{figure*}
\centering
\includegraphics[width=0.7\textwidth]{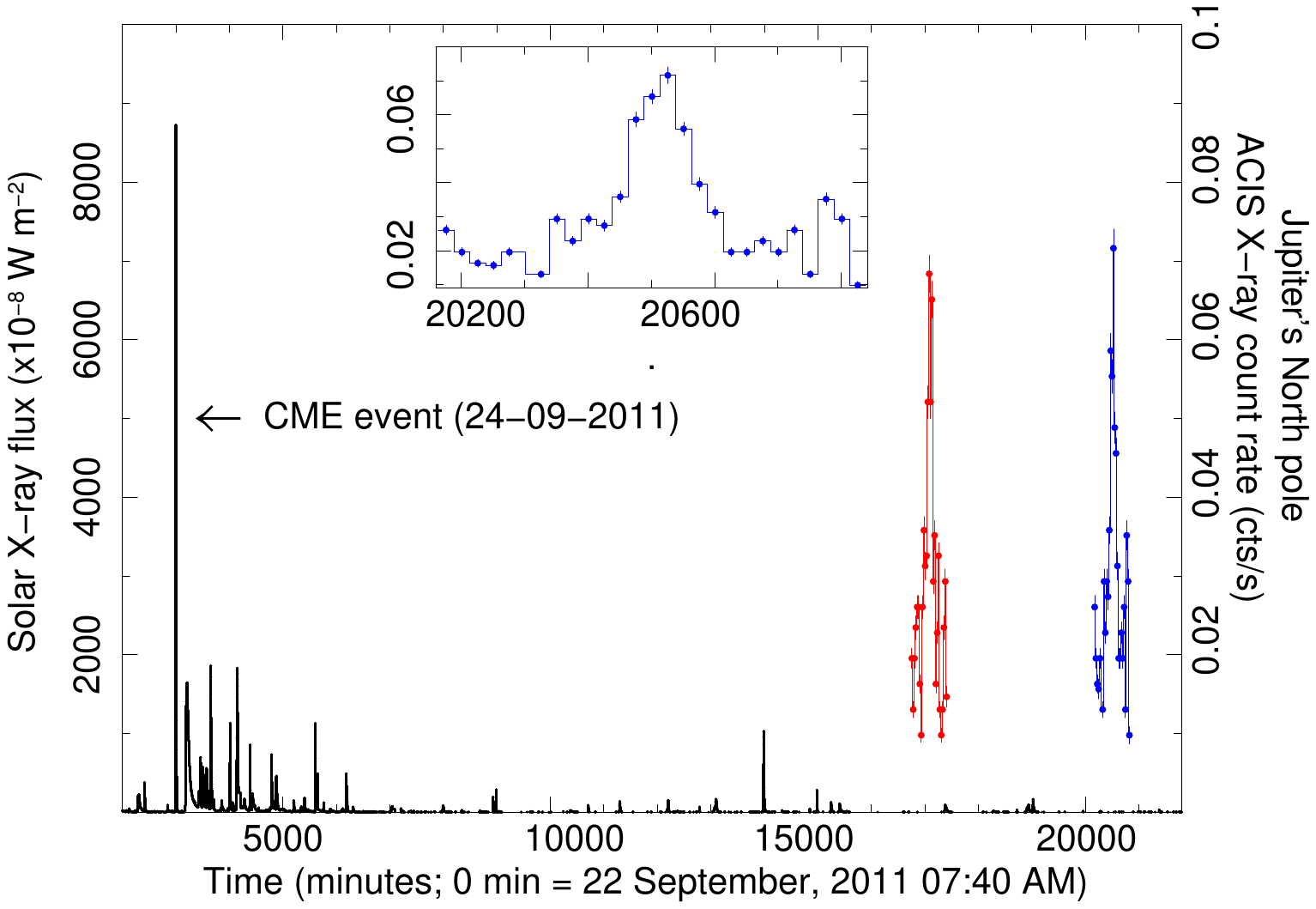}
\caption{The Solar X-ray activities, including a CME event, are shown along with the subsequent \textit{Chandra}/ACIS flaring lightcurves of Jupiter on 2nd October 2011 (red) and 4th October 2011 (blue). For clarity, the zoomed plot of the X-ray flare as observed on 4th October is shown in the inset.
}
\label{fig:6}
\end{figure*} 

\subsection{Spectral analysis}
\subsubsection{\textit{IUE} spectral analysis}
In Jupiter's \textit{IUE}/SWP data, we see clear variability in both the UV continuum and the emission lines. We may note that due to the non-imaging capabilities of \IUE{}, both disk and auroral regions will contribute to the observed flux in the UV spectra.
The top panel of Figure~\ref{fig:2} shows the UV data-to-model ratio spectra over 1150–1925~\AA\ for observations acquired during solar minimum (1987) and solar maximum (1991), obtained by fitting each spectrum with a power-law continuum, thereby highlighting systematic deviations associated with distinct spectral features.
Ratio plots clearly show that there are two Fe\textsc{ii} line complexes near 1600~\AA~during solar minima, which are missing from the other spectrum, raising the possibility that the UV spectral lines between the Solar minima and Solar maxima can be different. 
To track such changes over time, we extract continuum and line light curves by fitting each spectrum in \textsc{XSpec} using a $\chi^2$ minimisation routine. The spectra are modelled with a mix of narrow and broad Lorentzian components to capture the emission features and the underlying continuum. 
As seen in the top panel of Figure \ref{fig:3}, a single Lorentzian component does not fit the Ly$\alpha$ emission line and a strong line residual is observed around 1218 \AA~.
Hence, we fit two Lorentzian components to describe the particular line. While 1213 $\pm$ 1.6 \AA~ line is broad in all spectra (best fit values vary between 0.032-0.045 eV), 1218 $\pm$ 1.8 \AA~ line is consistently narrow (best fit values vary between 0.011-0.023 eV). \citet{2007A&A...463..761B} reported broad oxygen emission lines in \xmm{} spectra, indicating ion velocities of order 5000 km s$^{-1}$, consistent with precipitating oxygen ions undergoing charge exchange in Jupiter’s magnetosphere. In our case, the broad component corresponds to Doppler velocities of $\sim$900-1300 km s$^{-1}$, which may suggest a similar charge-exchange interaction of hydrogen ions. Therefore, both components of Ly$\alpha$ lines may originate from different geographical locations of Jupiter: the broader one may come from the aurora, where UV-emitting gas has strong dynamics, while the narrow component may originate from the disc, dominated by resonantly scattered or thermally broadened hydrogen emission. Throughout our work, we treat both line fluxes separately.
In addition to Ly$\alpha$, best-fit spectra require several Lorentzian functions depending on the number of emission lines. As an example, the best-fit model of the observation SWP23780 gives a reduced chi-square of $\chi^2$/d.o.f. = 164.67/155, with the best-fit spectra, model components, and residuals shown in the bottom panel of Figure \ref{fig:3} in 1150-1450~\AA~. We derive integrated fluxes for both these emission lines and the continuum (1370-1420~\AA) using the \textsc{cflux} convolution model in \textsc{XSpec}, which provides fluxes over chosen wavelength ranges along with their 1$\sigma$ errors. Each of the 51 spectra is fitted separately to obtain flux measurements at every pointing. The resulting fluxes are then corrected for Jupiter’s distance to ensure consistency across all observations.

\subsubsection{\textit{Chandra} spectral analysis}
The spectra near both auroral regions obtained with \textit{Chandra}/ACIS were modelled using the spectral fitting package \textsc{XSPEC v12}. Owing to the small number of spectral bins (10–15), the Kolmogorov–Smirnov (KS) statistic was preferred over the $\chi^{2}$ statistic for the goodness-of-fit analysis. Due to the red light leakage in soft X-ray bands and optical loading effect, the soft X-ray flux below 0.5 keV is contaminated beyond corrections in \chandra{}/ACIS spectra of Jupiter \citep{2005JGRA..110.1207E}. However, contamination is negligible in \chandra{}/HRC spectra. Therefore, we have excluded ACIS data analysis below 0.5 keV.
0.5-8 keV spectra obtained with \textit{Chandra}/ACIS were initially modelled using a bremsstrahlung continuum (\textit{bremss}) following earlier studies \citep{2007A&A...463..761B}. Significantly peaked residuals were observed, motivating the inclusion of the Gaussian emission line in the model. For example, for ObsID 12315, fitting with only the bremsstrahlung model gives a KS value of $-1.73$. Adding the first Gaussian at 0.7~keV improves the KS to $-2.58$, inclusion of the second Gaussian near 0.96–0.98~keV results in a total KS value of $-3.91$, indicating a substantially improved fit. A similar trend is seen for the 2011 \textit{Chandra}/ACIS obsID 12316: fitting with only the bremsstrahlung model gives a KS value of $-1.64$. Adding the first Gaussian at 0.7~keV improves the KS to $-2.52$; inclusion of the second Gaussian near 0.96–0.98~keV results in a total KS value of $-4.52$. The resultant fitting demonstrates that both line components are required to accurately reproduce the auroral spectra. 
The observed excess near 0.94--0.98~keV with broad line widths ($\sigma \approx 0.07$--0.14~keV) indicates the presence of a highly ionised element in Jupiter. We may note that a feature in a similar energy has been identified by \citet{2016JGRA..121.2274D} during the modelling of the hotspot quadrant spectra from \chandra{}. The line centre energy is consistent with the Ne \textsc{VIII} to Ne \textsc{IX} bound-bound transition.
The absence of Ne$^{8+}$ emission line from spectra previously reported in the literature \citep{2005JGRA..110.1207E, 2008JGRA..113.2202B, 2020JGRA..12527222D} suggests that the line is likely transient, possibly associated with enhanced solar radiation or magnetospheric conditions. The bremsstrahlung continuum temperatures are found to lie in the range $kT \approx 0.6$-1.3~keV. Best-fit parameters for both the spectra are provided in Table \ref{tab3}, and the fitted spectra along with the model components (shown by thin dotted lines) are shown in Figure \ref{fig:4}. Fluxes in the 0.5--10~keV band for different \textit{Chandra}/ACIS observations for the north and south auroral region are corrected for the Jupiter-Earth distance and reported in Table \ref{tab3}.

The top panels of Figure \ref{fig:4} show the fitted spectra with the 0.94~keV Gaussian feature excluded, revealing significant residuals around 0.9–1.0~keV. After including Gaussian line models near 0.94 keV, the residuals disappear. Fitted spectra after including additional Gaussian lines are shown in the bottom panels of Figure \ref{fig:4} along with residuals. For the Observation ID 12316, we assessed the significance of this newly identified line through Markov Chain Monte Carlo simulations of the spectral parameters from the joint spectral fits. When spectral modelling involves many free parameters (e.g., up to 7 in our best-fit model), $\chi^2$ minimisation alone is not always reliable for parameter estimation \citep{2012ApJ...755...88R}. Since X-ray spectral counts typically follow a Poisson distribution, we replace the conventional chi-squared statistic with \textsc{pgstat} in \textsc{XSpec}, which assumes Poisson-distributed source counts and Gaussian-distributed background counts. The profile likelihood of \textsc{pgstat} is derived in a manner similar to that of the C statistic. Using this statistic, we run MCMC chains of $10^6$ elements, discarding the first 50,000 as burn-in, and initialise the chains with random perturbations around the best-fit values. The proposal distribution at each Monte Carlo step is Gaussian, rescaled by a factor of 0.001. For rigorous testing, we adopt the Goodman-Weare algorithm with 60 walkers. Through this MCMC exercise, we find that the 0.94~keV line is statistically significant at least up to 3$\sigma$. Posterior contours corresponding to 1, 2 and 3$\sigma$ are shown in the top panel of Figure \ref{fig:5}.
We show the \textit{Chandra}/ACIS image extracted in the 0.8–1.2~keV energy range from the observations conducted on 02 October, 2011 (bottom panel of Fig.~\ref{fig:5}), highlighting the specific region associated with this emission. 
\label{sect:Obs}
\begin{table*}
\begin{center}
\centering
\caption{0.5-8 keV Best-fit parameters for the \textit{Chandra}/ACIS spectra of Jupiter’s northern and southern auroral region, with the width ($\sigma$) of the first Gaussian line fixed at 0.06 keV, and the listed fluxes include distance corrections.  detailed description of obs1 and obs2 is provided in Table \ref{Tab1}.}
\label{tab3}
\setlength{\tabcolsep}{6pt}
\small
\begin{tabular}{llccccccccc}
\hline
Component      & Parameter(unit) & obs1                    & obs2                                      \\
\hline
{\tt gaussian} & LineE(keV)      & $0.74^{+0.01}_{-0.01}$  & $0.72^{+0.02}_{-0.02}$   \\
               & norm(N)         & $7.37^{+1.31}_{-1.32}$  & $7.28^{+1.60}_{-1.71}$    \\
               & norm(S)         & $3.60^{+1.00}_{-0.98}$  & $2.04^{+0.97}_{-1.33}$   \\
               
{\tt gaussian} & LineE(keV)      & $0.98^{+0.02}_{-0.02}$  & $0.96^{+0.05}_{-0.07}$                         \\
               & sigma(keV)      & $0.07^{+0.02}_{-0.02}$  & $0.14^{+0.04}_{-0.04}$                         \\
               & norm(N)         & $3.74^{+1.18}_{-0.93}$  & $3.67^{+2.23}_{-1.53}$                         \\
               & norm(S)         & $2.34^{+0.73}_{-0.63}$  & $3.44^{+1.77}_{-1.23}$                        \\
               
{\tt bremss}   & kT(keV)         & 1.28(f)                 & $1.01^{+0.57}_{-0.28}$   \\
               & norm(N)         & $13.68^{+1.87}_{-1.95}$ & $9.54^{+5.90}_{-4.34}$ \\
               & norm(S)         & $5.27^{+1.45}_{-1.49}$  & $4.24^{+3.04}_{-1.94}$   \\
               
$\mathrm{flux_{0.5-10}}$(N) & ($10^{-14}$) & $4.34^{+0.19}_{-0.19}$ & $3.12^{+0.21}_{-0.21}$  \\
$\mathrm{flux_{0.5-10}}$(S) & ($10^{-14}$) & $1.94^{+0.18}_{-0.18}$ & $1.55^{+0.14}_{-0.14}$  \\
\hline
KS test  & (log)               & -3.91  & -4.52  \\
\hline

\end{tabular}
\end{center}
\end{table*}
\subsection{\textit{Chandra} Timing analysis}
 In addition to the spectral characterisation, the 2011 \textit{Chandra}/ACIS observations are analysed to study the temporal behaviour of Jupiter’s northern auroral X-ray emission. The datasets corresponding to Observation IDs 12315 and 12316, taken on 2nd and 4th October 2011, respectively, were analysed using identical data reduction and light-curve extraction procedures. The solar X-ray activity surrounding the \textit{Chandra}/ACIS X-ray flux of Jupiter on 2nd and 4th October 2011 is shown in Figure \ref{fig:6}. The background-subtracted light curve for the northern auroral region for Observation ID 12316 is shown in the inset of Figure \ref{fig:6}. These observations were selected due to limited data availability, and they were obtained 8 and 10 days after a major CME was reported, providing an opportunity to investigate whether enhanced solar activity influences Jupiter’s auroral X-ray flux. Examination of the light curves reveals a notable intensification in the auroral emission by a factor of $\sim$3 following the solar event. The temporal correlation between the increase in solar flux and the corresponding rise in Jupiter’s auroral X-ray flux is further illustrated in Figure \ref{fig:6}, highlighting the potential impact of solar activity on Jovian auroral processes.  In particular, enhancements in the 0.5–10 keV \textit{Chandra}/ACIS light curve appear approximately 8–10 days after a strong, recurring CME event associated with an M-class solar flare recorded between 24 and 26 September 2011. This temporal analysis complements the spectral results by providing insight into the short-term variability of the auroral region as described in the next section. 
 
To further investigate the temporal variability of Jupiter’s auroral X-ray emission, we analysed a total of 29 \textit{Chandra}/HRC observations spanning between 08 April 2014 and 15 September 2021(see Table~\ref{Tab2}). Background-subtracted light curves were extracted for all observations, and both flare and non-flare periods were examined to determine whether enhancements in Jupiter’s X-ray flux occur regularly or are linked to elevated solar activity. Two representative light curves, corresponding to 10 April 2014 and 15 April 2014, are shown in the bottom left and bottom right panels of Figure \ref{fig:7}, respectively. On the top right panel of Figure \ref{fig:7}, the \textit{Chandra}/HRC observations are shown for 10 and 15 April, along with the corresponding solar X-ray flux starting from 1st April, 2014, to provide the broader solar activity context leading up to and during the Jupiter observations. The enhanced X-ray flares can be observed from zoomed-in lightcurves shown in the bottom panels of Figure \ref{fig:7}. 
Since the hot spot is located near Jupiter's north pole and is fixed in longitude, it is visible for only about half of each Jovian rotation. As a result, one would expect the observed count rate to increase when the hot spot rotates into observer's line of sight and decrease when it rotates out of view, producing a broad modulation envelope as the hot spot traverses the visible hemisphere. Superposed on this underlying envelope, there may also be occasional quasi-periodic flaring events that contribute additional short-term variability.

\section{Correlation of Jupiter’s UV and X-ray Emissions with Solar Activity}
\label{sect:result}

\subsection{\textit{IUE} data comparison with solar activity}
\begin{figure}
    \centering
    \begin{subfigure}[t]{0.48\textwidth}
        \centering
        \includegraphics[width=\textwidth, angle=0]{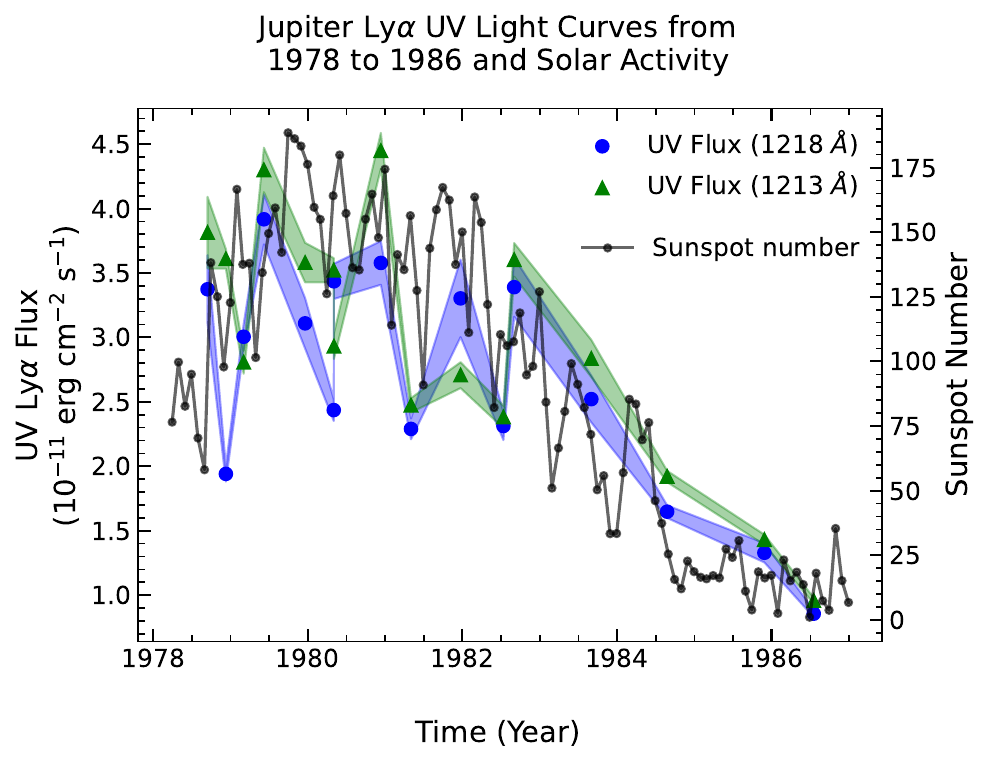}
    \end{subfigure}
    \hfill
    \begin{subfigure}[t]{0.48\textwidth}
        \centering
        \includegraphics[width=\textwidth, angle=0]{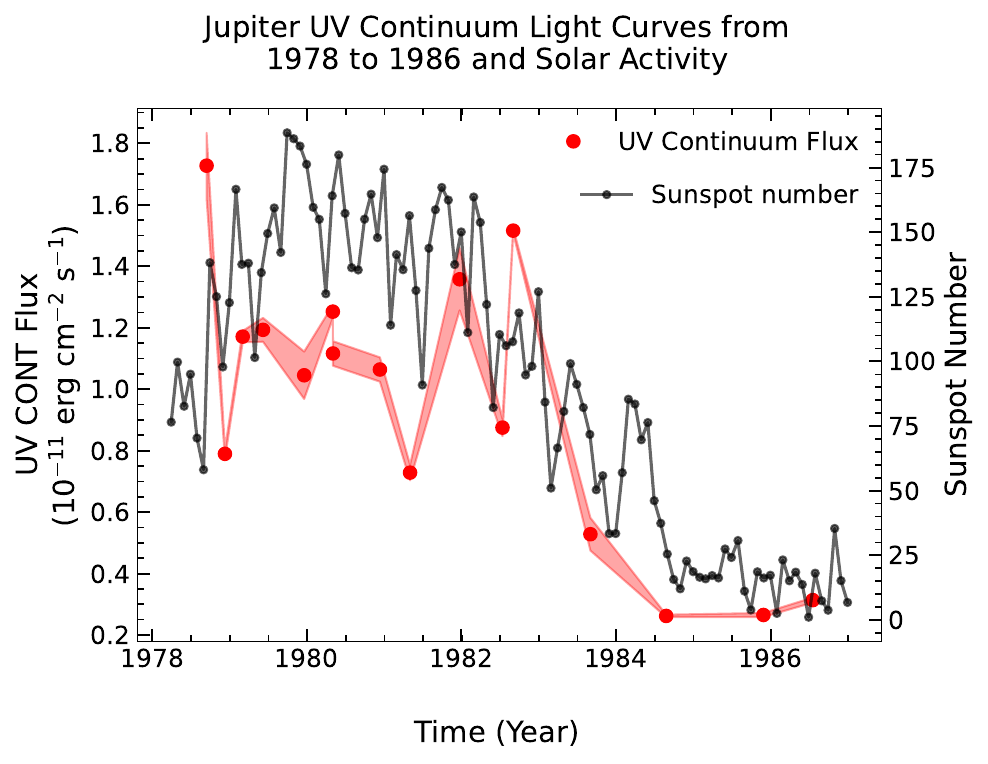}
    \end{subfigure}

    \caption{UV Ly$\alpha$ and continuum (1300-1400~\AA) flux variation of \textit{IUE}/SWP data from 1978 to 1986. UV flux is plotted alongside sunspot number data from the same period.}
    \label{fig:8}
\end{figure}

\begin{figure}
    \centering
    \begin{subfigure}[t]{0.48\textwidth}
        \centering
        \includegraphics[width=\textwidth, angle=0]{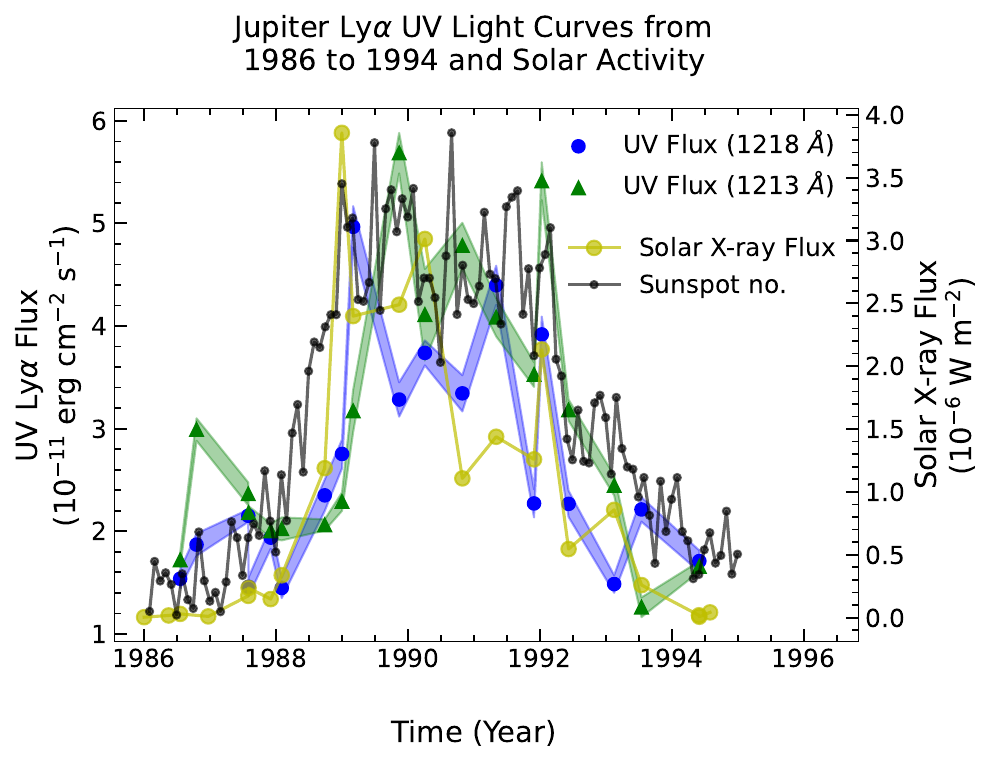}
    \end{subfigure}
    \hfill
    \begin{subfigure}[t]{0.48\textwidth}
        \centering
        \includegraphics[width=\textwidth, angle=0]{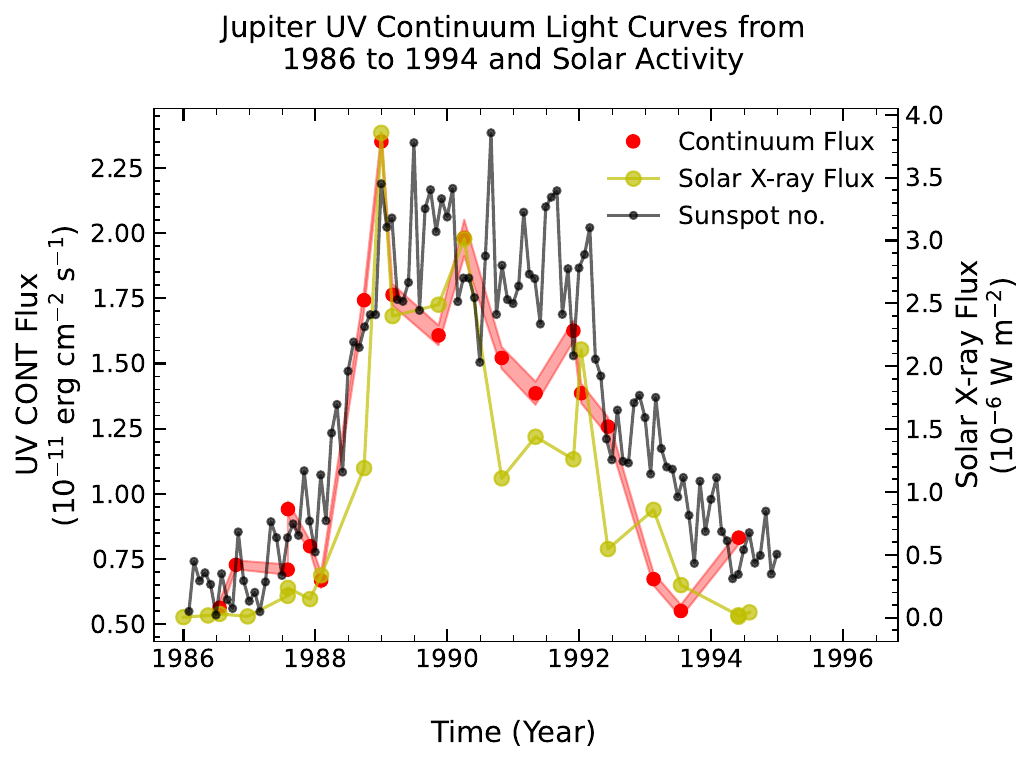}
    \end{subfigure}

    \caption{UV Ly$\alpha$ and continuum (1300-1400~\AA) flux variation of \textit{IUE}/SWP data from 1986 to 1994. UV flux is plotted alongside solar X-ray flux and sunspot number data from the same period. The sunspot numbers are normalised with respect to solar X-ray flux data.}
    \label{fig:9}
\end{figure}

\begin{figure}
    \centering
    \begin{subfigure}[t]{0.42\textwidth}
        \centering
        \includegraphics[width=0.7\textwidth, angle=270]{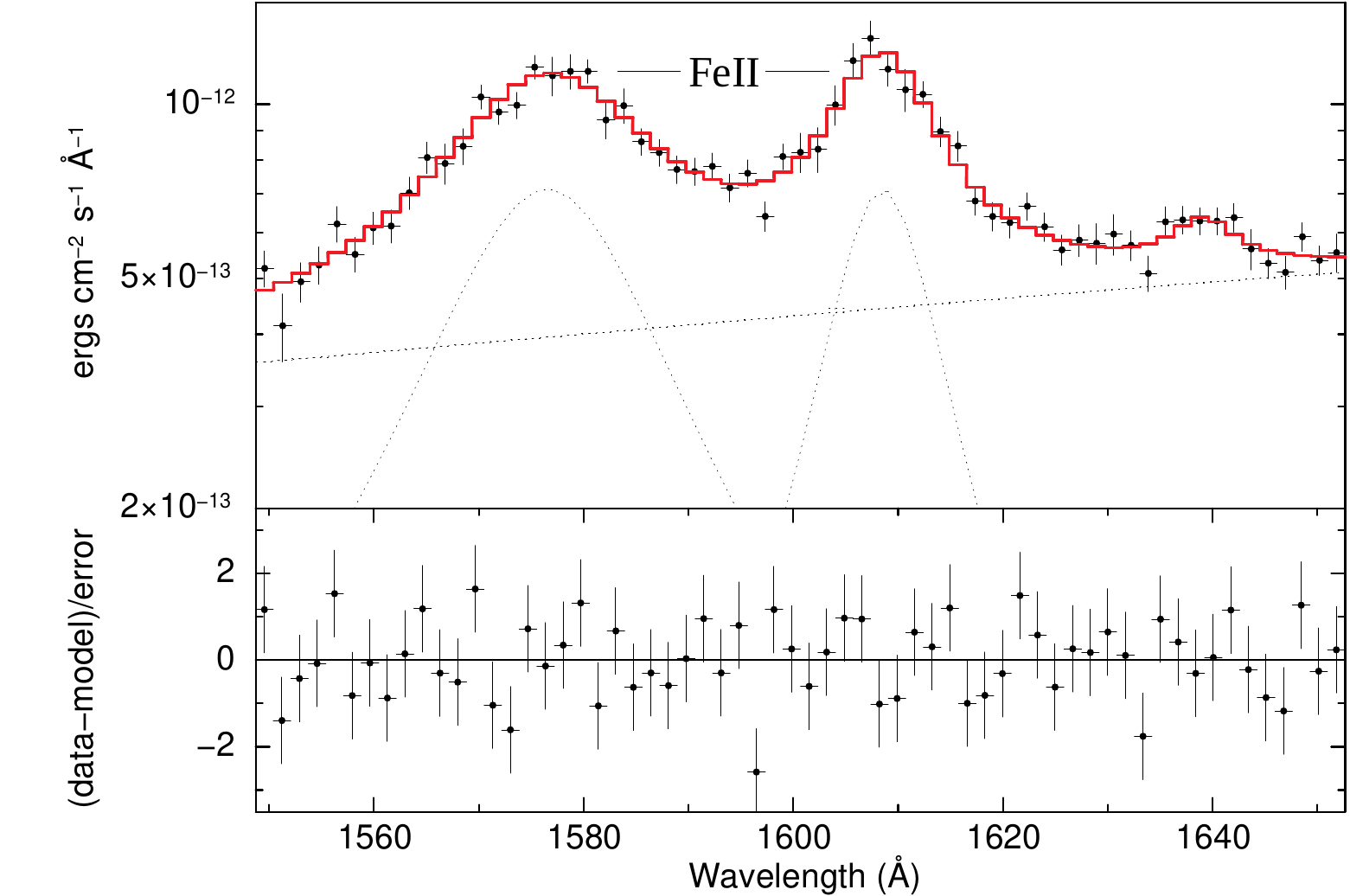}
    \end{subfigure}
    \hfill
    \begin{subfigure}[t]{0.42\textwidth}
        \centering
        \includegraphics[width=0.7\textwidth, angle=270]{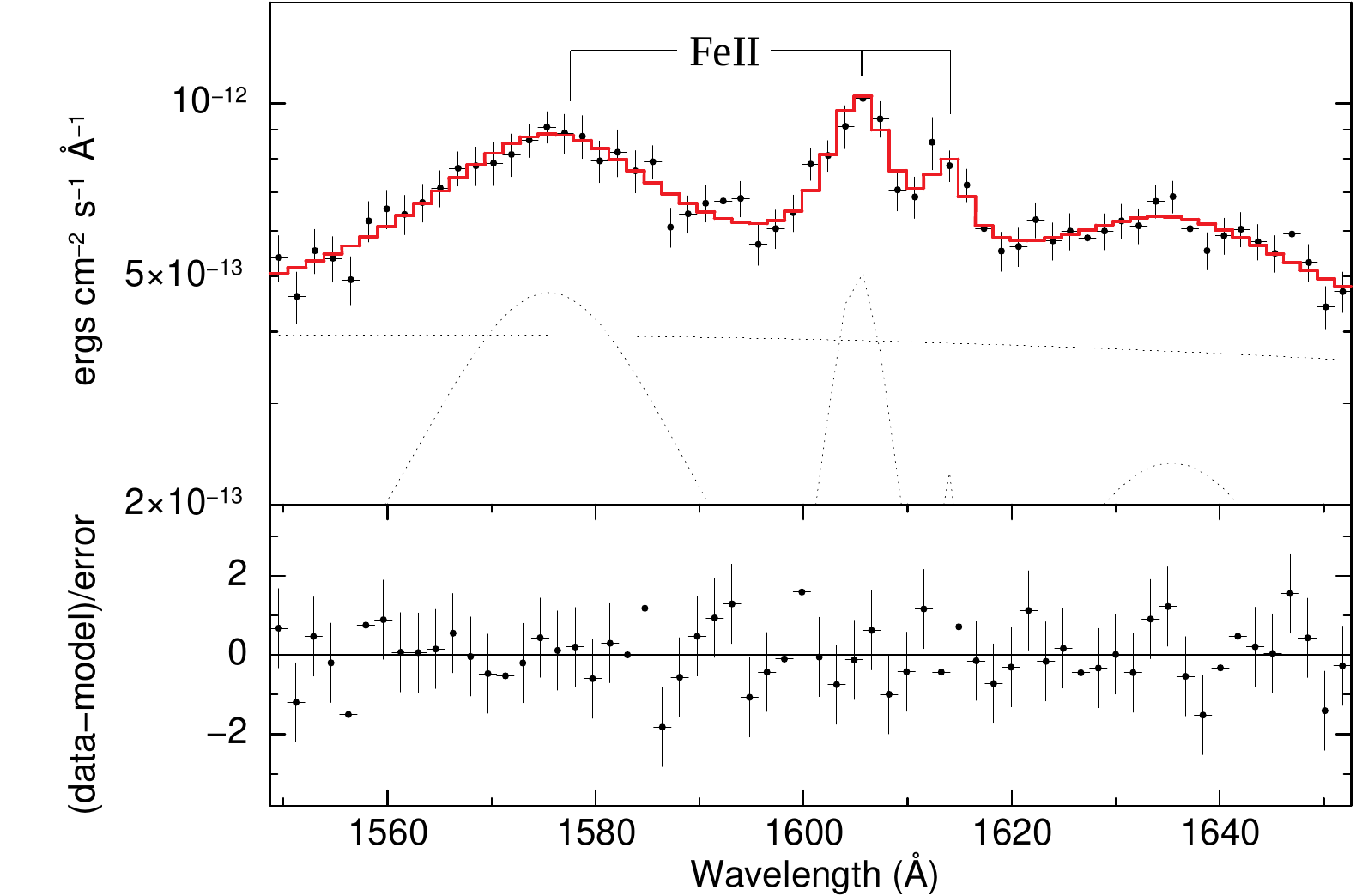}
    \end{subfigure}

    \caption{Spectral fitting in the range of 1550-1650~\AA. Top: The best-fit ($\chi^2$/d.o.f. = 50.61/50) UV spectrum is shown for the Observation ID \textsc{SWP31444}. Bottom: Best-fit ($\chi^2$/d.o.f. = 35.54/51) spectrum for the Observation ID SWP20739. The best-fit model (red) consists of continuum and emission line features, and the residual of the fitting is shown in the bottom panel.}
    \label{fig:10}
\end{figure}

\begin{figure}
    \centering
    \includegraphics[width=0.3\textwidth, angle=270]{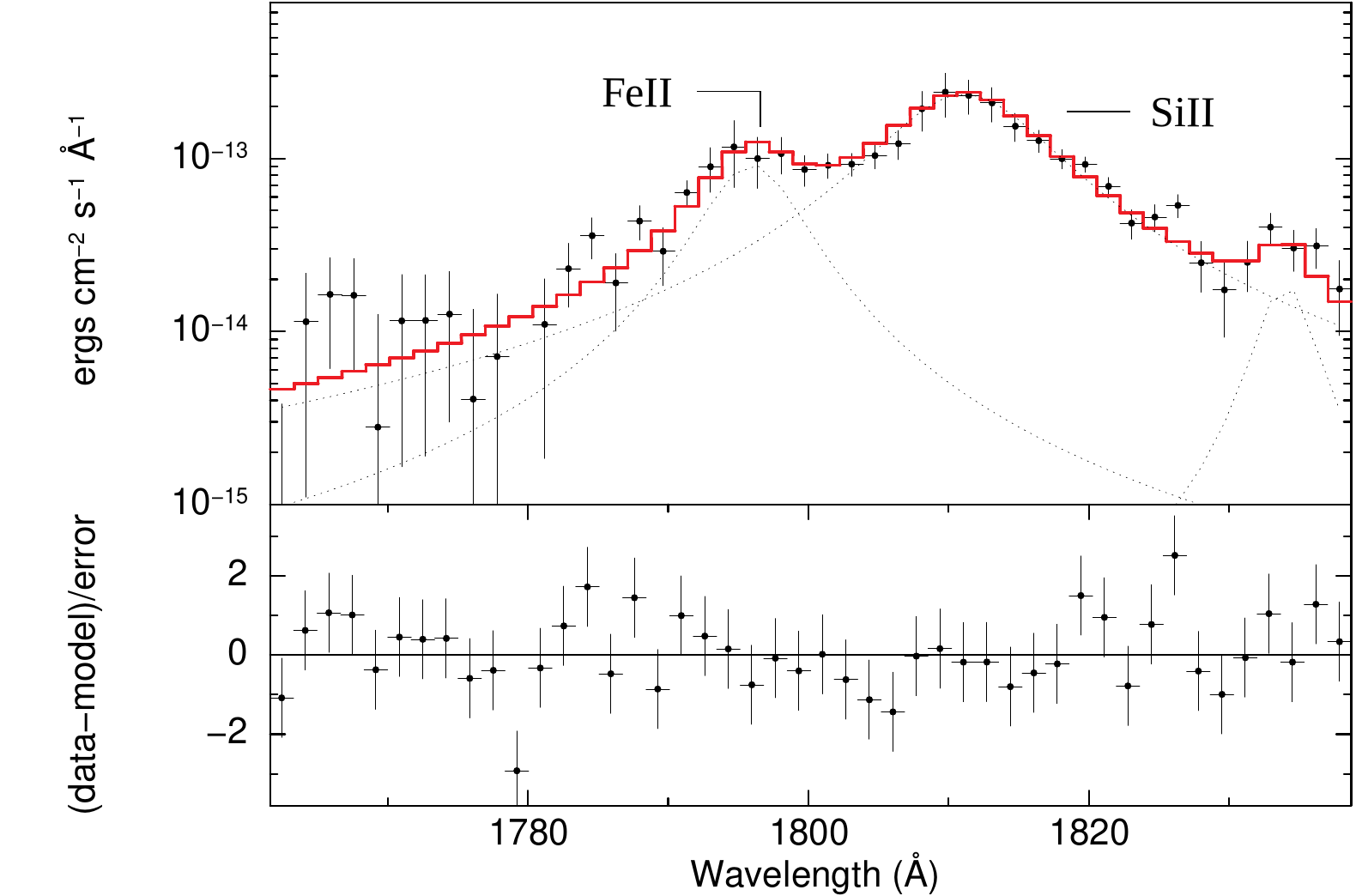}
    \caption{The best-fit ($\chi^2$/d.o.f. = 41.42/37) UV spectrum in 1760-1840~\AA~range is shown for the Observation ID SWP25688. The best-fit model (red) consists of emission line features, and the residual of the fitting is shown in the bottom panel.}
    \label{fig:11}
\end{figure}

\begin{figure*}
    \centering
    \includegraphics[scale=0.95]{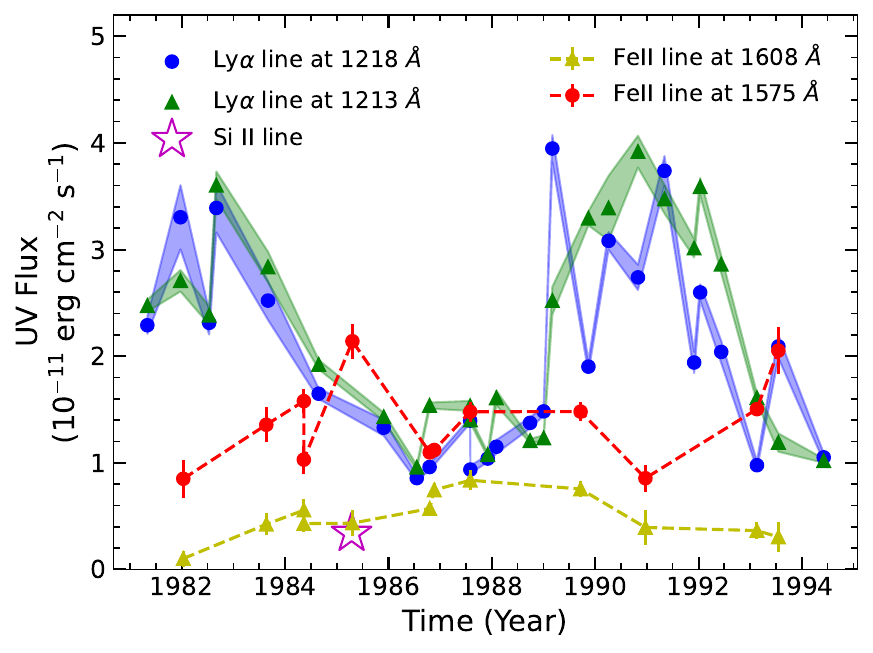}
    \caption{UV Ly$\alpha$ and Fe\textsc{ii} line flux variation of \textit{IUE}/SWP data from 1981 to 1994. Fluxes for both the Fe\textsc{ii} lines are fairly constant as compared to the significant variation for the Ly$\alpha$ lines. The magenta star denotes the observation time of Si\textsc{ii} line detection and its flux value.}
    \label{fig:12}
\end{figure*}

We plot the \IUE\ UV light curves for the Ly$\alpha$ line emissions and 1300-1400~\AA~continuum emission for the year 1978-1986 (21st solar cycle) in the left and right panels of Figure \ref{fig:8} and the same are shown in the left and right panels of Figure \ref{fig:9} for the year 1986-1996 (22nd solar cycle), which highlights the long-term variation of Jupiter's flux. We compare these fluctuations with simultaneous sunspot number data obtained from the \textsc{silso} database.
During the 1986–1996 period, X-ray flux data from the Sun, measured by the GOES-06 satellite, are also shown in Figure \ref{fig:9}. 
We have calculated the Pearson correlation coefficients between the Ly$\alpha$ flux and the solar flux, as well as between the continuum flux and the solar X-ray flux, to investigate their correlation. For Ly$\alpha$ at 1218~\AA, we obtain correlation coefficient $r = 0.678$ with a $p$-value of 0.002 and for Ly$\alpha$ at 1213~\AA~$r = 0.551$ with $p = 0.018$. While for the continuum emission $r = 0.89$ and $p = 7.37\times10^{-7}$ is calculated. These strong correlations indicate that increases in solar flux are reflected in Jupiter’s ultraviolet emission.
In several cases, particularly at the time of solar minima, we have also observed two significant peaks around 1575~\AA~and 1608~\AA. These features are likely due to stacking of multiple, unresolved Fe\textsc{ii} lines (at 1575.81, 1575.42, 1575.02~\AA~and at  1607.98, 1608.29, 1608.45, 1608.53, 1609.03, 1609.03, 1609.1~\AA). The spectral fitting of these two cases is shown in Figure \ref{fig:10}. 
We have detected Fe\textsc{ii} line complexes in 12 \IUE\ spectra, spanning between 1982 and 1994.
Interestingly, we have detected the Si\textsc{ii} line at 1812~\AA~during observation ID SWP25688 with a $p$-value of $2\times10^{-13}$ corresponding to a significance of $7.2\sigma$. The line is shown in Figure \ref{fig:11}. 
We calculated fluxes for both Fe\textsc{ii} lines during 12 observations. We compare Fe\textsc{ii} line flux variabilities with Ly$\alpha$ lines and show them in Figure \ref{fig:12}. The fluxes of both lines are fairly constant during rising, peak or descending phases of Ly$\alpha$ line variabilities (which follow X-ray flux variabilities during solar cycle) as shown in Figure \ref{fig:12}. Such a non-correlation indicates that Fe\textsc{ii} may not originate from solar interactions with Jupiter's atmosphere; rather, activities of Io may influence such origin, which we explore in the discussion section. Another possibility is the presence of iron in solar wind or magnetosphere, which may be photoionized by solar radiation. The line flux corresponding to Si\textsc{ii} is denoted as a magenta star in Figure \ref{fig:12}.
 
\subsection{HRC data comparison with solar activity}
\begin{figure}
  \centering
   \begin{subfigure}{0.48\textwidth}
    \includegraphics[width=\textwidth]{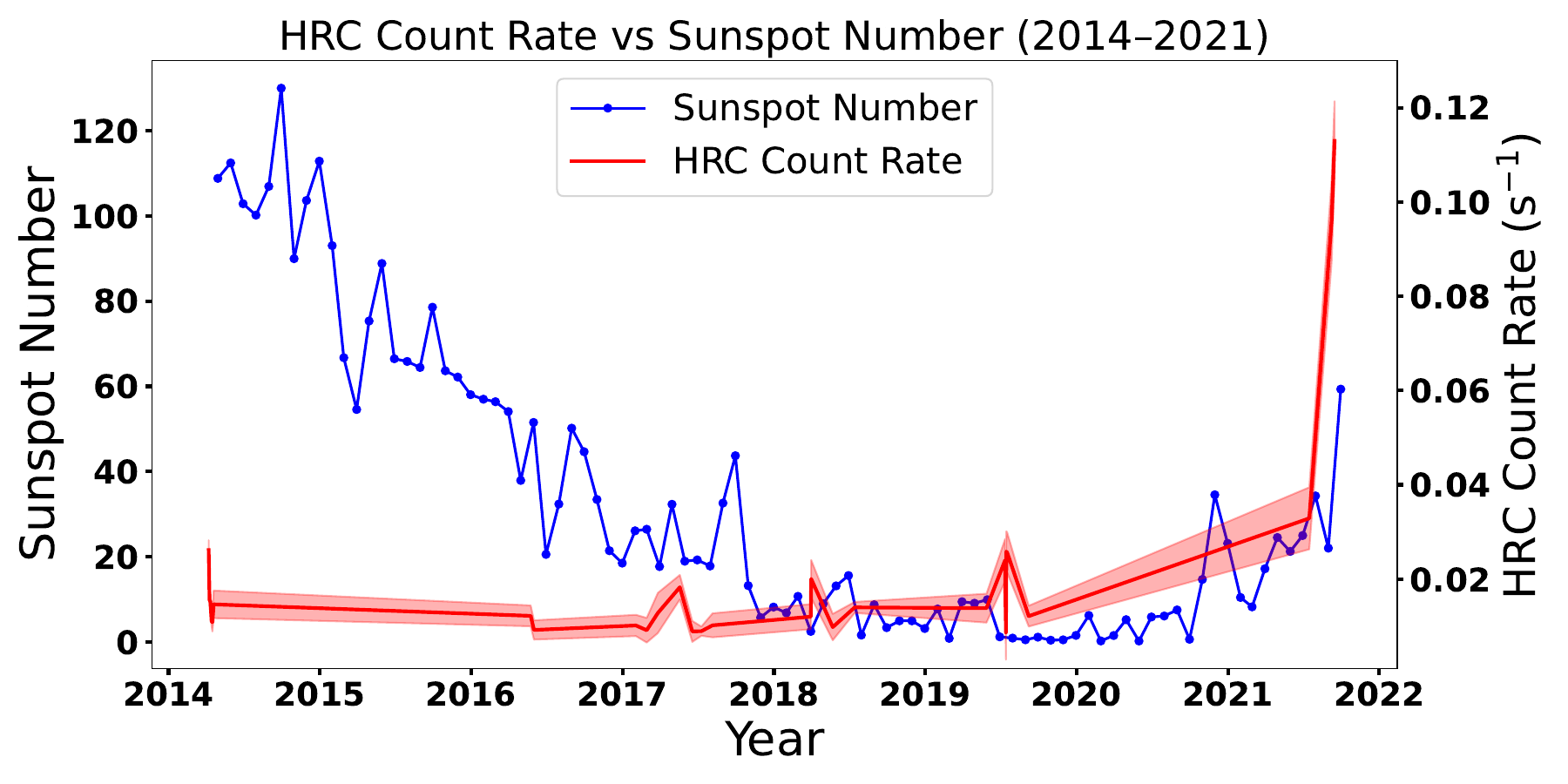}
    \label{fig:sub1}
  \end{subfigure}%
  \hfill
  \begin{subfigure}{0.48\textwidth}
    \includegraphics[width=\textwidth]{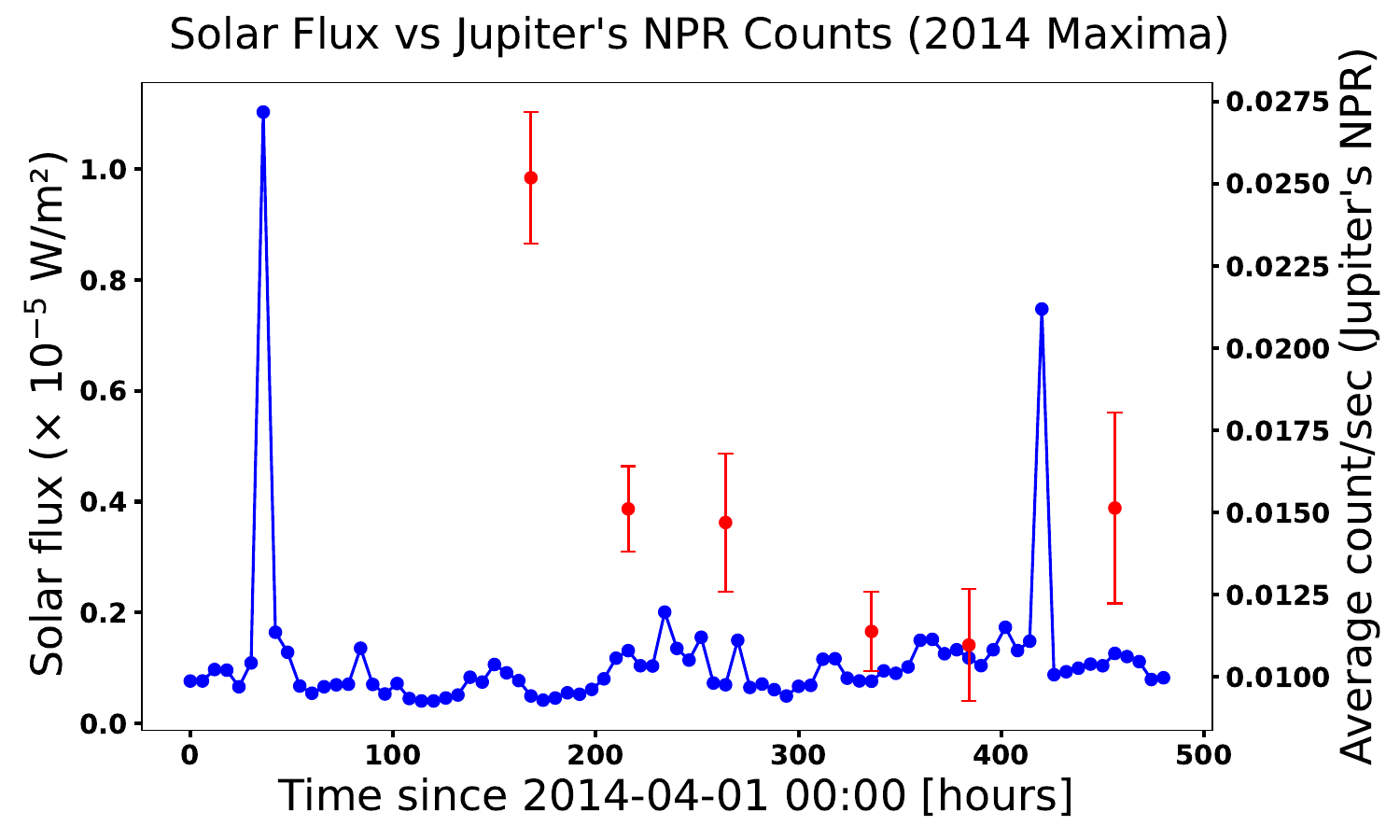}
    \label{fig:sub1}
  \end{subfigure}%
  \hfill
  \begin{subfigure}{0.48\textwidth}
    \includegraphics[width=\textwidth]{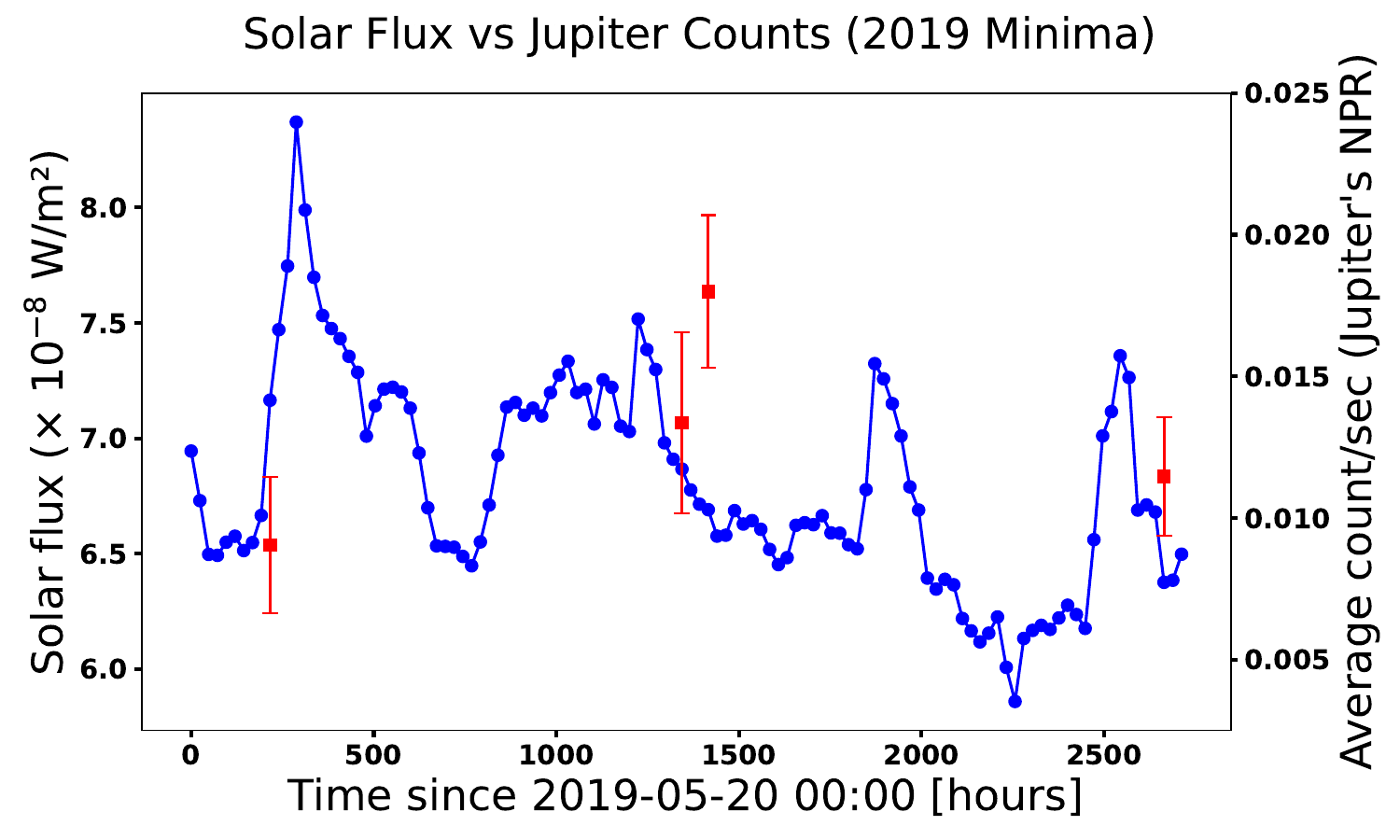}
    \label{fig:sub2}
  \end{subfigure}

  \caption{Top panel shows long-term (2014-2021) comparison of X-ray count rate (in log scale; after distance correction) obtained from each \textit{Chandra}/HRC observations with sunspot numbers. The distance-corrected count rate values are given in Table \ref{Tab2}. The middle and bottom panels show zoomed-in segments of the HRC light curve during solar maxima (2014) and solar minima (2019), respectively (shown by red circles), and corresponding solar X-ray flux values (shown in blue solid line). Neither long-term sunspot numbers nor short-term X-ray flux correlates with \chandra{}/HRC X-ray count rate from Jupiter.} 
  \label{fig:13}
\end{figure}

We examine the variability of Jupiter’s auroral emission in relation to solar activity (Figure \ref{fig:13}) for the \textit{Chandra}/HRC observations. The top panel of Figure \ref{fig:13} shows the comparison of average \textit{Chandra}/HRC-derived count rates with the monthly-averaged sunspot number, thereby placing the auroral variability within the broader framework of solar cycle evolution. No apparent long-term correlation is observed between Jovian X-ray intensity and solar X-ray flux, consistent with earlier work.  
The middle and bottom panels of Figure \ref{fig:13} show the zoomed segments for comparison during solar maxima (averaged over a 6-hour period during April 2014) and solar minima (May-September 2019). These datasets highlight the auroral response to varying levels of solar flux across different epochs. In either epoch, no significant correlation is observed, despite the limited number of data points.


\section{Discussion and Conclusions}
\label{sect:conclusion}
In this work, we present the solar activity-induced variability in Jupiter's ultraviolet and X-ray emissions using \textit{IUE} (1978–1996) and \textit{Chandra} (2011–2021) observations over multiple solar cycles. The UV analysis reveals that Jupiter's Ly$\alpha$ and continuum emissions strongly correlate with solar X-ray flux and sunspot numbers across two solar cycles, indicating a dominant solar irradiation origin, while ionised Fe\textsc{ii} emissions show no such correlation, suggesting a magnetospheric or internally driven source. In the X-ray regime, significant auroral flares are detected 7–15 days after major CMEs, along with a $\geq 3\sigma$ detection of highly ionised Ne$^{8+}$ emission near Jupiter's north pole, providing strong evidence for CME-driven auroral excitation and large-scale magnetospheric response to solar events.

\subsection{On correlation of UV continuum and Ly$\alpha$ line fluxes with solar activity}
Analysing 51 UV spectroscopic observations and comparing with the simultaneous solar activity parameters like X-ray fluxes and Sunspot numbers, we found that UV Ly$\alpha$ line and continuum fluxes from Jupiter’s atmosphere evolve significantly: during solar maxima to minima or vice versa, the UV Ly$\alpha$ line flux changes by a factor of $\sim$3.5 (as observed from top panels of Figure \ref{fig:8}) while the same during UV continuum changes by a factor of $\sim$4-5 (as observed from bottom panels of Figure \ref{fig:8}).
Figure \ref{fig:8} and Figure \ref{fig:9} show moderate to strong correlation (Pearson’s correlation coefficient ~0.68-0.91) with the evolution of solar X-ray fluxes as well as sunspot numbers during the rise-peak-decline phase of the first two solar cycles; Jupiter’s polar Lyman-$\alpha$ (1215.67 \AA) emission and far-UV continuum are highly sensitive to variations in solar extreme ultraviolet (EUV) and UV (UV) irradiance. This sensitivity arises because the upper atmosphere and exosphere of Jupiter contain a significant population of neutral hydrogen atoms, which act as resonant scatterers for incoming solar Ly$\alpha$ photons. The scattering process dominates the baseline Ly$\alpha$ brightness across the polar regions and is modulated directly by changes in the solar UV flux \citep{1980ApJ...241L.179C, 1998JGR...10320083G}
Solar EUV and X-ray photons play a crucial role in photochemistry in the Jovian upper atmosphere. High-energy UV and X-ray photons ionise and dissociate molecular hydrogen into atomic hydrogen, populating the exosphere and thus increasing the column density of atomic hydrogen for resonance scattering. Therefore, during times of peak solar activity marked by elevated sunspot numbers and observed EUV and X-ray fluxes, photodissociation becomes maximally efficient, and the density of scattering H atoms becomes very high. These scattering H atoms increase Ly$\alpha$ emissions and far-UV continuum in the auroral areas. The continuum emission arises primarily from scattered solar photons and weak molecular bands of H$_2$.
The years-long changes in Ly$\alpha$ brightness correlated with the solar cycle were studied by \citet{1980ApJ...241L.179C} and \citet{1993JGR....9818793G} using \IUE\ and \textit{Copernicus} observations, leading to the discovery that polar Ly$\alpha$ fluxes increase during maxima and diminish during minima of the solar cycles. Results from our work are also being confirmed by earlier findings by \citet{2000RvGeo..38..295B} and \citet{{2016Icar..268..215G}}.

\subsection{On non-correlation of solar activities with ionised UV lines}
Our results show that fluxes due to ionised UV lines like Fe\textsc{ii} at 1608 \AA~and 1575 \AA\ (Figure \ref{fig:10}) show no correlation with the solar activities studied over 12 years, as observed from Figure \ref{fig:12}. Ionised metal lines, such as Fe\textsc{ii} at 1575 \AA~and 1608 \AA, in Jupiter’s UV spectra exhibit a fundamentally different origin than the Ly$\alpha$ and far-UV continuum. Unlike Hydrogen lines, which are dominated by solar-driven resonant scattering, these metal lines cannot be efficiently excited by solar UV photons alone because the cross-sections for resonance scattering of Fe\textsc{ii} are small, and the solar flux at the relevant wavelengths is weak \citep{2006JGRA..111.4202G, 2019JGRA..124.8298G}. Instead, observations and modelling suggest that Fe\textsc{ii} and other heavy ion emissions are primarily generated by interactions of precipitating magnetospheric ions with the upper atmosphere \citep{2024SSRv..220...82Y}. These ions can be accelerated within Jupiter’s magnetosphere via processes such as centrifugal interchange instability, or magnetotail reconnection events, producing collisional excitation of $Fe^+$ ions in the atmosphere of Jupiter \citep{2003JGRA..108.1366G, 2015GeoRL..42.1662K}.
A particularly important source of such ions is the Io plasma torus. Io, with its intense volcanic activity, injects large quantities of neutral and ionised sulphur, oxygen, and trace metals, including iron, into Jupiter’s magnetosphere \citep{2011JGRA..116.5209B}. These ions become entrained in the corotating plasma and are transported along magnetic field lines toward the auroral regions, where precipitation leads to collisional excitation of metal ions and emission in the UV. The flux of such Fe\textsc{ii} lines is therefore strongly modulated by magnetospheric dynamics, for example, rotational modulation of the magnetosphere, episodic plasma injections, interchange instability events, or Io-related plasma variations- rather than by the long-term solar cycle.
\subsection{On the origin of Si$\textsc{II}$ line}
The detection of Si\textsc{ii} (1812 \AA) in Jupiter’s spectra during solar minima, shown in Figure \ref{fig:11} and Figure \ref{fig:12} is particularly intriguing because it appears only when the solar-driven UV background is low, suggesting that the line is either inherently weak or masked by stronger solar-scattered emissions during periods of high solar activity. This behaviour indicates that the observed Si\textsc{ii} line is likely dominated by local or magnetospheric processes rather than by direct solar excitation.
One plausible explanation is enhanced visibility of weak, locally generated emissions during solar minima. When the solar UV flux is low, the background Ly$\alpha$ and far-UV continuum decrease, reducing the competing signal from resonance scattering. This allows intrinsically faint auroral emissions, such as Si\textsc{ii} produced by particle precipitation or low-density plasma processes, to become detectable \citep{2006JGRA..111.4202G, 2019JGRA..124.8298G}.
A second possibility involves episodic local particle acceleration or upwelling, which can selectively populate the excited states of $Si^+$ ions. Processes such as field-aligned acceleration of electrons or heavy ions, auroral potential drops, or localised reconnection events may energise ions in the polar upper atmosphere, leading to collisional excitation of Si\textsc{ii} \citep{2003JGRA..108.1366G, 2015GeoRL..42.1662K}. Such events may be sporadic and independent of the long-term solar cycle, explaining why the line appears only intermittently.
A third mechanism is magnetospheric reconfiguration, such as plasma injections, centrifugal interchange events, or sudden rearrangements of the magnetotail. These processes can temporarily increase the flux of magnetospheric metal ions along field lines into the polar atmosphere, producing a transient enhancement in Si\textsc{ii} emission \citep{2011JGRA..116.5209B, art2020}.
To differentiate among these possibilities, time-tagged spectral observations combined with in-situ magnetospheric measurements (e.g., from \textit{JUNO} or future missions) or simultaneous remote sensing of auroral morphology would be critical. 

\subsection{On CME-induced auroral activities}

On two occasions in 2011 and 2014, we reported two X-ray flares observed with \chandra{}/ACIS (Figure \ref{fig:6}) and two flares from \chandra{}/HRC (Figure \ref{fig:7}), 7-15 days after major CMEs were detected. Due to the scarcity of Jupiter's X-ray observations after CMEs, the beginning and the end of such flares cannot be determined. Using same \chandra{}/ACIS observations in 2011, \citet{2016JGRA..121.2274D} found that due to an interplanetary coronal mass ejection (ICME), Jupiter’s X-ray aurora intensified by nearly a factor of eight, accompanied by rapid changes in spatial, spectral, and temporal behaviour, including two hotspot periodicities from distinct ions: 26 min from sulphur ions and 12 min from a carbon/sulphur and oxygen. Their analysis showed that sulphur-dominated emissions map to mid-magnetospheric field lines (50-70 R$_J$), while mixed oxygen–sulphur–carbon emissions map to more distant or open field lines, suggesting a possible solar-wind contribution. Together with coincident radio bursts, they observed that both the auroral enhancement and the behaviour of the polar X-ray hot spot were driven by ICME-induced magnetospheric compression and/or pulsed dayside reconnection. It has been proposed that enhanced solar-wind disturbances associated with CMEs can compress Jupiter's magnetosphere and trigger dayside magnetopause reconnection, leading to plasma injections and magnetospheric reconfiguration \citep{nichols2006}.
To test whether observed X-ray flares in the lightcurve and highly-ionised emission line in the spectra (bottom panels of Figure \ref{fig:7}) can be induced by CME events, we calculate the closeby coronal mass arrival time on the Jovian atmosphere.

On 2011 September 24, multiple coronal mass ejections (CMEs) were detected with notably high velocities\footnote{\url{https://spaceref.com/status-report/joint-usafnoaa-report-of-solar-and-geophysical-activity-24-september-2011/}}. The first major event occurred around 09:48 UT, associated with an X1.9 flare at approximately 09:21 UT, followed by a second eruption at about 12:48 UT, connected to an M7.1 flare at 12:33 UT.
During both the X-ray CME launch and the subsequent \chandra{} observations (after a week), the Sun-Jupiter-Chandra geometry was highly favorable, with \chandra{} viewing Jupiter at a very small phase angle (observer and solar directions differing by only $\sim$ 6-7$^\circ$), providing an almost fully illuminated view of Jupiter's northern auroral region and enabling a direct investigation of the Jovian response to solar activity.
The CME speed measured is roughly 1915–2089 km/s at 20 R$_s$ (Solar Radius) according to the \textit{SOHO}/LASCO catalogue\footnote{\url{https://cdaw.gsfc.nasa.gov/CME_list/}} with the kinetic energy of 3 $\times$ 10$^{32}$ ergs. From the same location, the second CME speed is measured around 972-788 km/s with the kinetic energy of 1.4 $\times$ 10$^{31}$ ergs \citep{2015yCat..18020053A}.

Considering the above speed estimations, we assume 972 km/s and 2089 km/s as the lower and upper bounds for CME speed at which they leave the Sun. Using Keplerian propagation from J2000 orbital elements\footnote{\url{https://ssd.jpl.nasa.gov/horizons/app.html##/}}, the Sun-Jupiter average distance is estimated to be 741942506.90 km. (averaging between 24 September and 3rd October), Then the CME will reach roughly between 4.5 and 9 days, considering the upper and lower bounds of speed. 
Therefore, the CME encounter on the Jovian atmosphere is expected to occur between 29 September and 03 October, 2011. 
However, following Drag Based Model (DBM) for the propagation of CME beyond 1 AU, proposed by \citet{2014ApJS..213...21V}, we expect a moderate drag by solar wind such that CME accelaration, $a = \gamma (V_{CME}-V_{wind})|V_{CME}-V_{wind}|$, where $V_{CME}$ and $V_{wind}$ are CME speed and ambient solar wind speed respectively while $\gamma$ is drag parameter. With the typical $\gamma$ value of 1-5 $\times$ 10$^{-8}$ km$^{-1}$ for a fast CME \citep{vr08,2010A&A...512A..43V,2015ApJS..218...32Z}, we may expect a further delay on arrival of CMEs by a couple of days. Therefore, CME-induced activities are expected to be detected 7-11 days (between 01 and 05 October, 2011) from the origin of CMEs. 

This is supported by our results: the significant detection of extremely ionised Ne$^{8+}$ line signatures in X-ray spectra, along with enhanced Bremsstrahlung emission (Figure \ref{fig:4} and top panel of Figure \ref{fig:5}), and strong flarings in X-ray lightcurves observed with \chandra{} on 2nd October and 4th October, 2011 (Figure \ref{fig:6}).
Following the above arguments, we may note that transient solar events (M-class flares, ICME/CME-driven shocks) can compress Jupiter's magnetosphere, trigger dayside reconnection or injections, and increase particle acceleration, producing episodic X-ray brightening as observed in the top right and bottom panels of Figure \ref{fig:7}. As observed from Figure \ref{fig:7}, \chandra{}/HRC count rate increases by 4-8 times on two occasions: 6-8 days and 12-15 days after the CME, consistent with solar wind/CME propagation time scales of the order of a week and interaction and subsequent magnetospheric response as discussed earlier.

\subsection{On the origin of Ne$^{8+}$ emission feature}
Ne$^{8+}$ is a high charge state ion, formed at coronal temperatures of a few million degrees Kelvin. Such ions are abundant in the solar wind, especially during CMEs or high-speed streams. The solar wind consistently carries a fraction of highly charged neon ions, detectable in situ by spacecraft like \textit{ACE}, \textit{Ulysses}, and \textit{Solar Orbiter}. Two plausible mechanisms may explain the origin of the Ne$^{8+}$ line in the auroral region: (a) Solar-wind/CME material injection and charge exchange (SWCX) \citep{2021PASJ...73..504A}: when a CME carrying high-charge-state heavy ions reaches Jupiter, those ions undergo charge exchange with exospheric neutrals, producing characteristic line emission. (b) As proposed in \citet{2016JGRA..121.2274D}, magnetospheric acceleration of solar-derived heavy ions: CME-induced compression/reconnection accelerates heavy ions (from solar wind or magnetospheric reservoirs) onto polar field lines where they produce ion-line emission by charge exchange or direct collisional excitation. The observed time delay matches expected transit times for fast CMEs with moderate values of drag parameter ($\gamma \simeq$1-5 $\times$ 10$^{-8}$ km$^{-1}$) to Jupiter and the results of dedicated \chandra{} campaigns that linked strong X-ray enhancements to solar storms. Together, these lines of evidence point to direct CME-induced auroral excitation on Jupiter.

\section*{Acknowledgements}
We are grateful to the referee for constructive comments, and we gratefully acknowledge the detailed discussions and constructive suggestions provided by Dr William Dunn and Bryn Parry, which significantly improved the quality of the manuscript. M.T. is grateful to the University Grants Commission of India (UGC-Grant No. 231620114372) and the Ministry of Science and Technology, Government of India, for financial support of the research work carried out at the Indian Institute of Technology, Hyderabad. AB acknowledges DST, Government of India, for the award of INSPIRE fellowship (IF230384) and the research facility of the Department of Physics, IIT Hyderabad.

\section*{Data Availability}
This paper is based on publicly available observations obtained with \chandra\ and \IUE\ with instruments and contributions directly funded by ESA Member States and the USA. The high-level data underlying this article are extracted through standard processing from raw data stored in HEASARC and MAST public archives.

\label{lastpage}

\end{document}